\documentclass[11pt,aps,prd,showpacs,amsmath,amssymb,floatfix,nofootinbib,superscriptaddress]{revtex4-2}

\usepackage{graphicx}  % needed for figures
\usepackage{multirow}
\usepackage[dvipsnames]{xcolor}
\usepackage[colorlinks=true,allcolors=BlueViolet]{hyperref}
\usepackage{latexsym}
\usepackage{amsfonts}
\usepackage{amsmath,amssymb}
\usepackage{slashed}
\usepackage{color}
\usepackage{orcidlink}
\usepackage{supertabular} 
\usepackage{epsfig}
\usepackage{lipsum}

\newcommand{\bqn}{{\boldsymbol q}_n}
\newcommand{\bq}{\boldsymbol q}
\newcommand{\phase}{\text{cot}\,\delta}

\newcommand{\itp}{\affiliation{CAS Key Laboratory of Theoretical Physics, Institute of Theoretical Physics, \\Chinese Academy of Sciences, Beijing 100190, China}}

\newcommand{\ucas}{\affiliation{School of Physical Sciences, University of Chinese Academy of Sciences, Beijing 100049, China}}
\newcommand{\imp}{\affiliation{Institute of Modern Physics, Chinese Academy of Sciences, Lanzhou, 730000, China}}
\newcommand{\ific}{\affiliation{Instituto de F\'isica Corpuscular (IFIC), CSIC-UV,
46980 Paterna, Valencia, Spain}}

\newcommand{\scnt}{\affiliation{Southern Center for Nuclear-Science Theory (SCNT), Institute of Modern Physics,\\ Chinese Academy of Sciences, Huizhou 516000, China}}

\newcommand{\PKU}{\affiliation{ School of Physics, Peking University, Beijing 100871, China}}

\newcommand{\PKHEP}{\affiliation{Center for High Energy Physics, Peking University, Beijing 100871, China}}

\newcommand{\ICQM}{\affiliation{Collaborative Innovation Center of Quantum Matter, Beijing 100871, China}}

\begin{document}

\title{{Low-energy $DD$ scattering in lattice QCD}
\\[1.5em]
\includegraphics[scale=0.3]{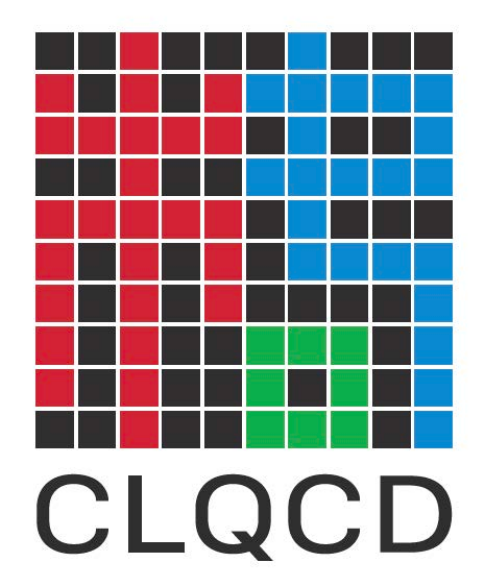}}

\author{Pan-Pan~Shi\orcidlink{0000-0003-2057-9884}}\email{Panpan.Shi@ific.uv.es}
\ific\itp\scnt
\author{Feng-Kun~Guo\orcidlink{0000-0002-2919-2064}}\email{fkguo@itp.ac.cn}
\itp\ucas%\peng\scnt

\author{Chuan Liu\orcidlink{0000-0003-4497-6868}}
\PKU\PKHEP\ICQM

\author{Liuming~Liu\orcidlink{0000-0002-7988-0631}}\email{liuming@impcas.ac.cn}
\imp\ucas

\author{Peng Sun}
\imp\ucas

\author{Jia-Jun~Wu\orcidlink{0000-0003-4583-7691}}\email{wujiajun@ucas.ac.cn}
\ucas\scnt

\author{Hanyang~Xing}
\imp\ucas

% \date{\today}

\begin{abstract}
We present the first lattice QCD calculation of single-channel $DD$ scattering with quantum numbers $I(J^P)=1(0^+)$ and $0(1^-)$. The calculation is performed on the $2+1$ flavor Wilson-Clover ensembles with a lattice spacing $a\simeq 0.077$ fm and two different pion masses, $m_{\pi}\simeq207$ and $305$ MeV. The scattering parameters are determined using the L\"uscher's finite volume method. 
Our results indicate a weak repulsive interaction in the $1(0^+)$ channel and a slightly attractive interaction in the $0(1^-)$ channel. The $S$-wave isovector $DD$ scattering length and effective range, extrapolated to the physical pion mass, are {$(-0.25\pm0.08\pm 0.12)$~fm and $(-5.7\pm4.5\pm 1.7)$~fm}, respectively.  
\end{abstract}

\maketitle

\section{Introduction}

Quantum Chromodynamics (QCD) is the fundamental theory that describes the strong interaction in terms of the quark and gluon degrees of freedom. However, in the low-energy region (at the energy scale close to $\Lambda_{\text{QCD}}$), a perturbative expansion in powers of the strong coupling constant fails, and thus alternative methods are required to study the strong interaction at this scale. One such method is lattice QCD (LQCD), which allows for investigations of the strong interaction in the nonperturbative regime from first principles. In LQCD, hadron scattering information can be extracted using L\"uscher's finite-volume method~\cite{Luscher:1985dn,Luscher:1986pf,Luscher:1990ux} and its generalizations (for a review, see Ref.~\cite{Briceno:2017max}).
The method relates the energy spectrum of a two-particle system in the finite volume and the scattering information of the two particles in the continuum. Following this path, tremendous progress has been made with LQCD in understanding hadron spectra and interactions, see recent reviews in Refs.~\cite{Briceno:2017max,Mai:2021lwb,Bulava:2022ovd,Mai:2022eur}. 
As for scatterings of identical spinless mesons, $\pi\pi$ scattering has been extensively investigated by several LQCD groups for various isospin channels $I=0$, $1$ and $2$~\cite{CP-PACS:2004dtj,Beane:2007xs,NPLQCD:2011htk,Dudek:2012gj,Fu:2013ffa,Bulava:2016mks,Wu:2021xvz,Sasaki:2013vxa,Blanton:2021llb}, and $KK$ scattering in the $I=1$ channel has also been explored in Refs.~\cite{Beane:2007uh,Sasaki:2013vxa,Helmes:2017smr,Blanton:2021llb}.
Here we will focus on $DD$ scattering in the low-energy regime.

$DD$ scattering is related to the study of exotic states in the doubly charmed sector. In particular, the doubly charmed exotic meson $T_{cc}(3875)^+$, which couples strongly to the isoscalar $DD^*$ in the $S$ wave, was observed in the $D^0D^0\pi^+$ invariant mass distribution by LHCb~\cite{LHCb:2021vvq,LHCb:2021auc}. Information on $DD$ scattering can provide insights into the $DD$ final state interactions that influences predictions on the $T_{cc}^+$ width~\cite{Yan:2021wdl, Fleming:2021wmk, Dai:2023mxm}.\footnote{For explorations on the influence of the analogous $D\bar D$ final state interaction on the open-charm decay width for $X(3872)\to D^0\bar D^0\pi$, see Refs.~\cite{Guo:2014hqa,Dai:2019hrf}.} 
Actually, the phenomenological calculation in Ref.~\cite{Yan:2021wdl} displayed a sizeable impact of the $DD$ final state interaction on the $T_{cc}$ width based on the one-boson-exchange model~\cite{Liu:2019stu}. 
In addition, the $DDK$ three-body bound state has been predicted in both continuum~\cite{MartinezTorres:2018zbl,Wu:2019vsy,Huang:2019qmw, Pan:2025xvq} and finite volume formalisms~\cite{Pang:2020pkl,Xiao:2024dyw}. So far, in the inclusive decay processes of $\Upsilon(1S)$ and $\Upsilon(2S)$~\cite{Belle:2020xca}, the Belle Collaboration has not found significant signals associated with this state in the $D^+D_s^{*+}$ invariant mass distribution. To further investigate the possible $DDK$ bound state, model-independent input on $DD$ two-body scattering is essential.

LQCD has been employed to study the interaction of the $DD^*$ system. The $DD^*$ scattering~\cite{Padmanath:2022cvl,Chen:2022vpo,Lyu:2023xro} and the $DD^*$-$D^*D^*$ coupled-channel scattering~\cite{Whyte:2024ihh} have been explored using the interpolating operators of $DD^*$ and both $DD^*$ and $D^*D^*$, respectively. Additionally, the quark mass dependence of $T_{cc}^+$ has been analyzed with $DD^*$ and $D^*D^*$ interpolating operators~\cite{Collins:2024sfi} (see also Ref.~\cite{Abolnikov:2024key}). While analyzing the LQCD data for the $DD^*$ system, it was noted that the one-pion exchange long-range force (the associated left-hand cut and a zero of the scattering amplitude) could significantly affect the extraction of scattering information~\cite{Du:2023hlu}. Several approaches have been proposed to address this issue~\cite{Meng:2023bmz,Raposo:2023oru,Bubna:2024izx,Hansen:2024ffk,Du:2024gzw}. One natural solution is to include the $DD\pi$ three-body operator in the $DD^*$ scattering process. {The formalism for the $DD\pi\to DD\pi$ three-body scattering can be utilized to analyze the lattice data, allowing the extraction of the $D^*D\to D^*D$ amplitude at energies where $D^*$ is treated as a $D\pi$ bound state or resonance. This method remains valid below the $DD\pi$ left-hand cut.} However, the lack of isovector $DD$ and isoscalar $DD\pi$ scattering information hinders further precise studies of the $T_{cc}^+$~\cite{Hansen:2024ffk}.

This work presents the first LQCD calculation of $DD$ scattering based on the $2+1$ flavor full QCD ensembles. The L\"uscher finite-volume method is utilized to connect the finite-volume energy levels to infinite-volume scattering parameters. We compute the energy levels with the $DD$ interpolating operators for both $I=1$ and $I=0$. The scattering length and effective range for both $S$- and $P$-wave channels are extracted. The calculations are performed at two different pion masses, $m_{\pi}\simeq 305$ MeV and $m_{\pi}\simeq207$ MeV. The pion mass dependence of the $S$-wave scattering length and effective range is explored.

This paper is organized as follows. In Sect.~\ref{Sec:Lattice}, details of the lattice setup and methodology are introduced. In Sect.~\ref{sec:operator}, we discuss the single- and two-meson operators as well as the method for the extraction of the energy levels from correlation functions. The formula, involving the  L\"uscher equation modified by the dispersion relation and the parameterization of phase shift, is discussed in Sect.~\ref{sec:lushcer_fromula}. The $S$- and $P$-wave scattering information for the isovector and isoscaclar $DD$ channels, including the scattering lengths, effective ranges, and the phase shifts, are {presented} in Sect.~\ref{sec:result}. We explore the pion mass dependence of the $S$-wave scattering length and effective range in Sect.~\ref{sec:pion_mass}. A conclusion is given in Sect.~\ref{sec:summary}. Finally, Appendix~\ref{Sec:dsp_rlt} contains details of the derivation of the L\"uscher equation with the modified dispersion relation.

\section{Lattice Setup}\label{Sec:Lattice}

\begin{table*}[tb]
	\centering
    \renewcommand\arraystretch{1.6}
 	\caption{Parameters of the ensembles, including the name, the lattice spacing $a$, the pion mass $m_{\pi}$, the volume size $L^3\times T$, the number of configurations $N_{\text{conf.}}$, {and the value of $m_{\pi}L$}. }
  \begin{ruledtabular}
  \begin{tabular}{lccccc}
	Ensembles	& $a$ [fm] & $m_{\pi}$ [MeV] & $(L/a)^3\times T/a$ & $N_{\text{conf.}}$ & {$m_{\pi}L$}\\
	\hline
    F32P30			&	$0.07746(18)$		& $305.2(13)$ & $32^3\times 96$ & 371  & {$3.83 (2)$}\\
	F48P30			&	$0.07746(18)$		& $305.4(9)$ & $48^3\times 96$ & 201 & {$5.75 (2)$} \\
    F32P21			&	$0.07746(18)$		& $207.9(22)$ & $32^3\times 64$ & 201 & {$2.61(3)$} \\
    F48P21			&	$0.07746(18)$		& $207.2(11)$ & $48^3\times 96$ & 223  & {$3.90(2)$}\\
	\end{tabular}
  \end{ruledtabular}
	\label{Tab:configuration}
\end{table*}

The results presented in this paper are based on the gauge configurations generated by the CLQCD Collaboration with $2+1$ dynamical quark flavors using the tadpole improved tree-level Symanzik gauge action and Clover fermion action ~\cite{CLQCD:2023sdb}. Numerous studies have been performed on these configurations, see, e.g., Refs.~\cite{Zhang:2021oja, Xing:2022ijm, Liu:2022gxf, Liu:2023feb, Yan:2024yuq, CLQCD:2024yyn, Meng:2024gpd}. In this work, we use four ensembles with the same lattice spacing $a = 0.07746$~fm and two different pion masses $m_{\pi}\simeq 305$~MeV and $m_{\pi}\simeq207$~MeV. Details of the ensembles are listed in Table~\ref{Tab:configuration}. Among these ensembles, F32P21/F48P21 and F32P30/F48P30 are two couples that share the same pion mass but have different volumes to obtain more kinematic points in the finite-volume spectra, rendering a more stable and precise determination of the scattering parameters. The valence charm quark action is the same as the light/strange quark action used in the ensembles, and the charm quark mass is tuned to reproduce the spin-averaged $1S$-charmonium mass, $m_{\text{av}}=(m_{\eta_c}+3m_{J/\psi})/4$.

The quark propagators are computed with the distillation quark smearing method~\cite{HadronSpectrum:2009krc}, which improves the precision and allows for the inclusion of many interpolating operators in the calculation of the correlation functions. The smearing operator is composed of a small number ($N_{\rm ev}$) of the eigenvectors associated with the $N_{\rm ev}$ lowest eigenvalues of the three-dimensional Laplacian defined in terms of the HYP(hypercubic)-smeared gauge field. The number of eigenvectors $N_{\rm ev}$ is 100 for all four ensembles. We also tried $N_{\rm ev}$=200 for the ensembles F48P21 and F48P30, and the resulting energies were consistent with {those obtained using} $N_{\rm ev}$=100. The statistical uncertainty is estimated by the bootstrap method with 4000 bootstrap samples. 

\section{Calculation of the energy levels}\label{sec:operator}

In LQCD, the finite-volume spectra are extracted from the two-point correlation functions
\begin{align}
{\cal C}_{ij}(t)=\sum_{t_s}\left\langle{\cal O}_{i}(t+t_s){\cal O}^{\dag}_{j}(t_s)\right\rangle,
\label{Eq:correlation}
\end{align}
where ${\cal O}_i$ and ${\cal O}_j$ are the interpolating operators carrying the quantum numbers of the system of interest. The source time $t_s$ spans all time slices to increase statistics.

\begin{table}[tb]
  \caption{\label{Tab:dispersion} Parameters for the dispersion relation of the $D$ meson. }
  \renewcommand{\arraystretch}{1.2}
  \begin{tabular*}{\columnwidth}{@{\extracolsep\fill}lcccc}
  \hline\hline 
    & F32P30 & F48P30   &  F32P21  & F48P21 \\[3pt]     
  \hline
       $m_D$ [MeV] & 1966.6(8)
       & 1966.0(6)  & 1903.9(16)   & 1900.3(5) \\[3pt]
       $Z$ & 0.926(7)  & 0.946(9) & 0.941(1)    & 0.976(6)   \\[3pt]
  \hline\hline
  \end{tabular*}
  \end{table}

\begin{figure*}
  \centering	
  \includegraphics[width=0.49\textwidth]{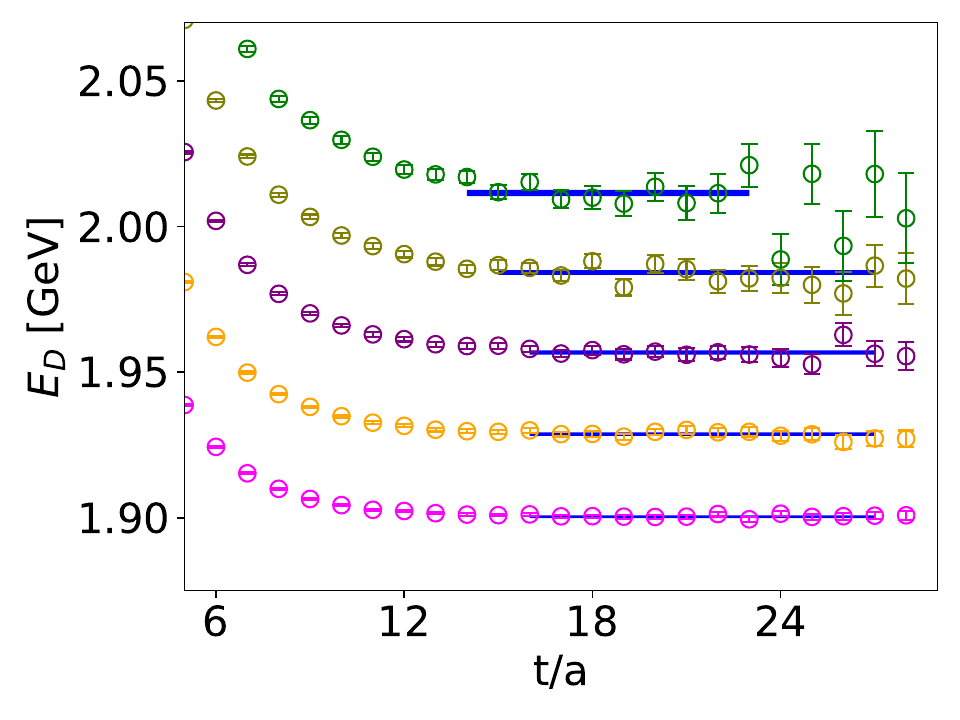}
  \hfill 
  \includegraphics[width=0.49\textwidth]{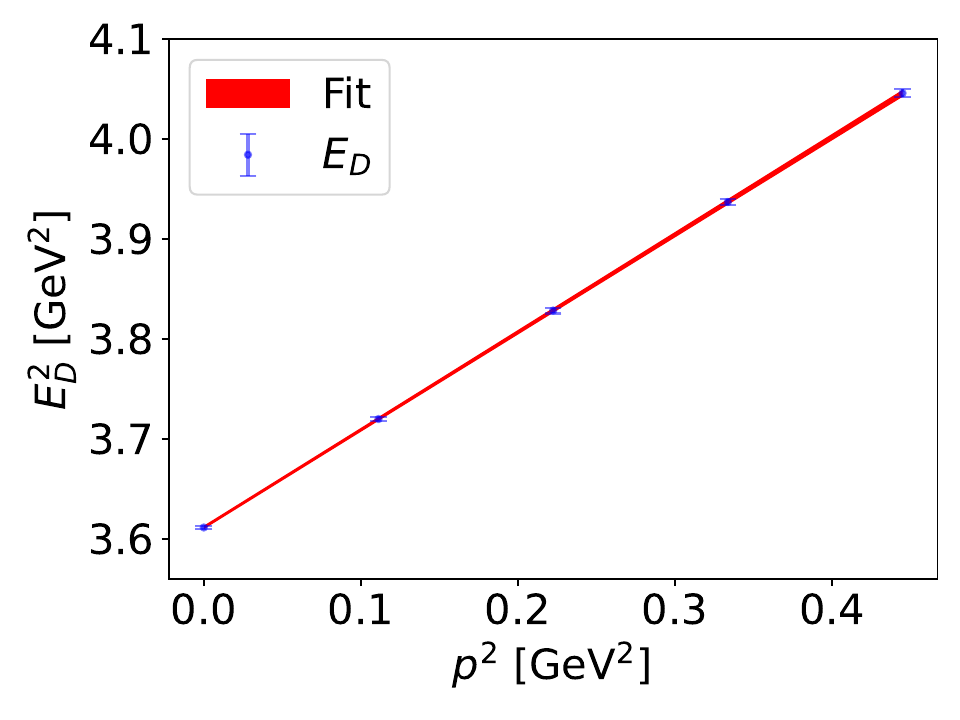}
  \caption{Effective energies and the dispersion relation of the $D$ meson for the ensemble F48P21. The left panel presents five energies of the $D$ meson at the five lowest momenta, and the right panel presents a fit of the dispersion relation to these energies. \label{Fig:D_mass}}
  \end{figure*}

The single $D$-meson operators are ${\cal O}_{D^+}(\boldsymbol{x}, t)=\bar d\gamma_5 c(\boldsymbol{x}, t)$ and ${\cal O}_{D^0}(\boldsymbol{x}, t)=\bar u\gamma_5 c(\boldsymbol{x}, t)$ for the charged and neutral $D$, respectively. The momentum projected operator is defined as $\mathcal{O}(\boldsymbol{p}, t) = \sum_{\boldsymbol{x}} e^{-i\boldsymbol{p}\cdot\boldsymbol{x}} \mathcal{O}(\boldsymbol{x}, t)$.  The dispersion relation of the $D$ meson $E^2 = m_D^2 + Zp^2$, with $p\equiv \sqrt{\boldsymbol{p}^2}$,  is investigated by calculating the single-particle energy at the five lowest momenta on lattice: $\boldsymbol{p} =$ $(0,0,0)$, $(0,0,1)$, $(0,1,1)$, $(1,1,1)$ and $(0,0,2)$ (in units of $2\pi/L$). 
The parameters $m_D$ and $Z$, listed in Table~\ref{Tab:dispersion}, are determined by fitting the five energies to the dispersion relation. 
As an example, the five energies and the dispersion relation for the ensemble F48P21 are shown in Fig.~\ref{Fig:D_mass}.  In the continuum limit, $Z$ should be unity. However, due to lattice artifacts, it may take a different value. The original L\"uscher's formula assumes the continuum dispersion relation. In order to account for the effects of the deviation  from the continuum dispersion relation, we modify the L\"uscher's formula as elaborated in Appendix~\ref{Sec:dsp_rlt}. 

The interacting energy levels of two particles are extracted from the correlation functions of the two-particle operators. The interpolating operators for the isovector and isoscalar $DD$ systems are given by
\begin{align}
{\cal O}_{DD,I=1}^{\boldsymbol P,\Lambda,\mu }&=\frac{1}{\sqrt 2}\sum_{\boldsymbol p_1, \boldsymbol p_2}{\cal C}(\boldsymbol P, \Lambda,\mu,\boldsymbol p_1, \boldsymbol p_2)\left({\cal O}_{D^+}(\boldsymbol p_1){\cal O}_{D^0}(\boldsymbol p_2)+{\cal O}_{D^0}(\boldsymbol p_1){\cal O}_{D^+}(\boldsymbol p_2)\right),\\
{\cal O}_{DD,I=0}^{\boldsymbol P,\Lambda,\mu }&=\frac{1}{\sqrt 2}\sum_{\boldsymbol p_1, \boldsymbol p_2}{\cal C}(\boldsymbol P, \Lambda,\mu,\boldsymbol p_1, \boldsymbol p_2)\left(-{\cal O}_{D^+}(\boldsymbol p_1){\cal O}_{D^0}(\boldsymbol p_2)+{\cal O}_{D^0}(\boldsymbol p_1){\cal O}_{D^+}(\boldsymbol p_2)\right),
\end{align}
where $\boldsymbol P=\boldsymbol p_1 + \boldsymbol p_2$ is the total momentum, $\Lambda$ denotes the irreducible representation (irrep) of the octahedral group and its subgroups, which are the spatial rotational symmetry groups on lattice, $\mu$ denotes the irrep row, and ${\cal C}(\boldsymbol P, \Lambda,\boldsymbol p_1, \boldsymbol p_2)$'s are the Clebsch-Gordan coefficients. In this work, we only compute the energies in the rest frame ($\boldsymbol{P}=\boldsymbol{0}$) and the momenta of the single particles up to $|\boldsymbol p_1| = |\boldsymbol p_2| = 2$. For $I=1$, the operators reside in the $A_1^+$, $E^+$ and $T_2^+$ irreps, while for $I=0$, the operators are in the $T_1^-$, $T_2^-$, and $A_2^-$ irreps. The relevant Clebsch-Gordan coefficients are provided in Appendix A of Ref.~\cite{Dudek:2012gj}. 

{We do not include compact tetraquark-like interpolating operators in the present calculation. Local color-neutral four-quark operators can be Fierz-rearranged into products of quark bilinears in general~\cite{Weinberg:2013cfa}. For lattice diquark-antidiquark interpolators, this relation and its limitations for smeared and momentum-projected operators are discussed explicitly in Ref.~\cite{Padmanath:2015era}. Thus such operators do not introduce an additional two-hadron channel below the $DD$ threshold, although they may have different overlap with a compact component of the finite-volume eigenstates. In the near-threshold region analyzed here, the lowest relevant two-meson channel is the $DD$ channel, while other meson-meson components with the same valence content correspond to higher thresholds or excited states. We therefore expect the low-lying levels entering the $DD$ scattering analysis to be dominated by the $DD$ interpolators included in our basis.
This expectation is supported by the lattice study of Ref.~\cite{Cheung:2017tnt}, where in doubly charmed channels, including the $DD_s$ sector, adding compact tetraquark-like operators to a meson-meson basis did not significantly change the extracted low-lying finite-volume spectra. A dedicated calculation including such operators in the present $DD$ setup would be useful to quantify this possible source of systematic uncertainty. 
} 

\begin{figure*}
\centering	
\includegraphics[width=0.48\textwidth]{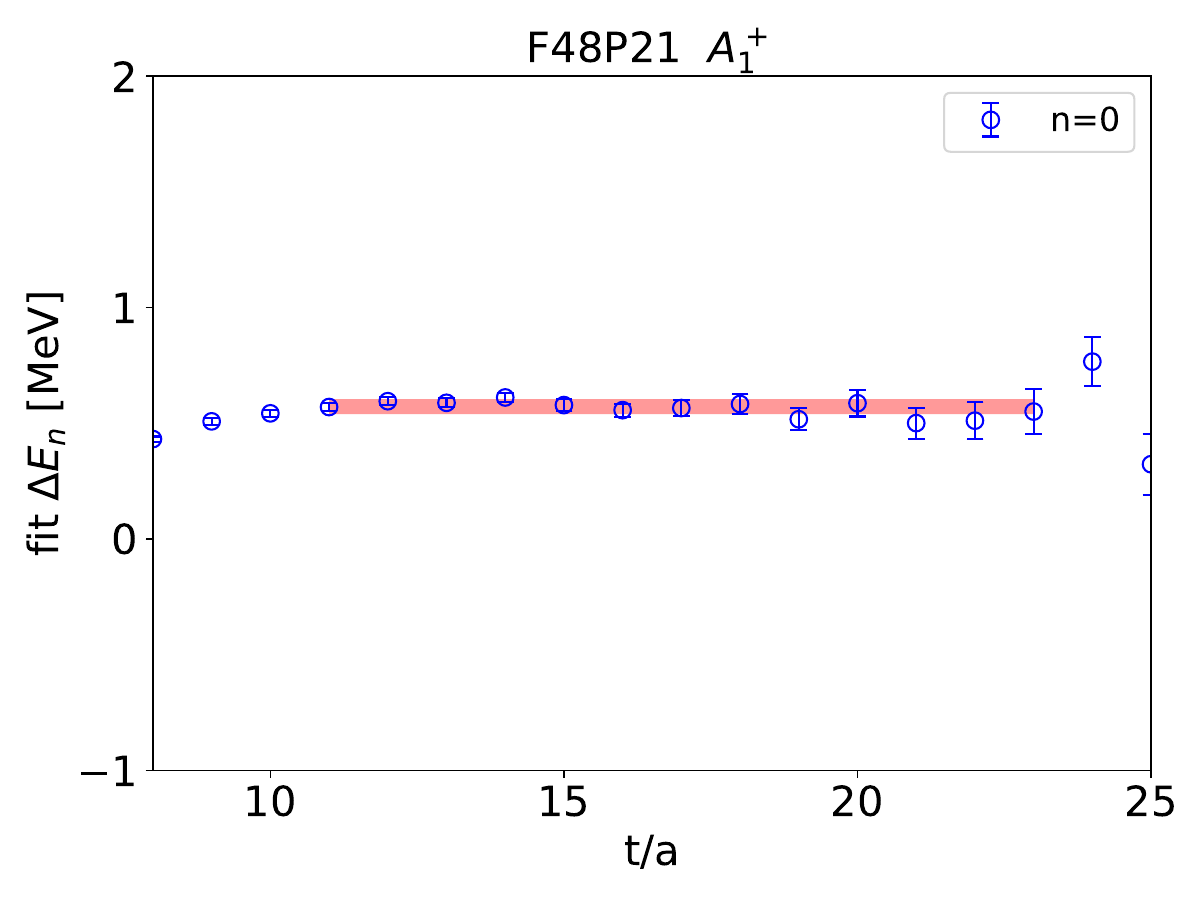} \hfill
\includegraphics[width=0.48\textwidth]{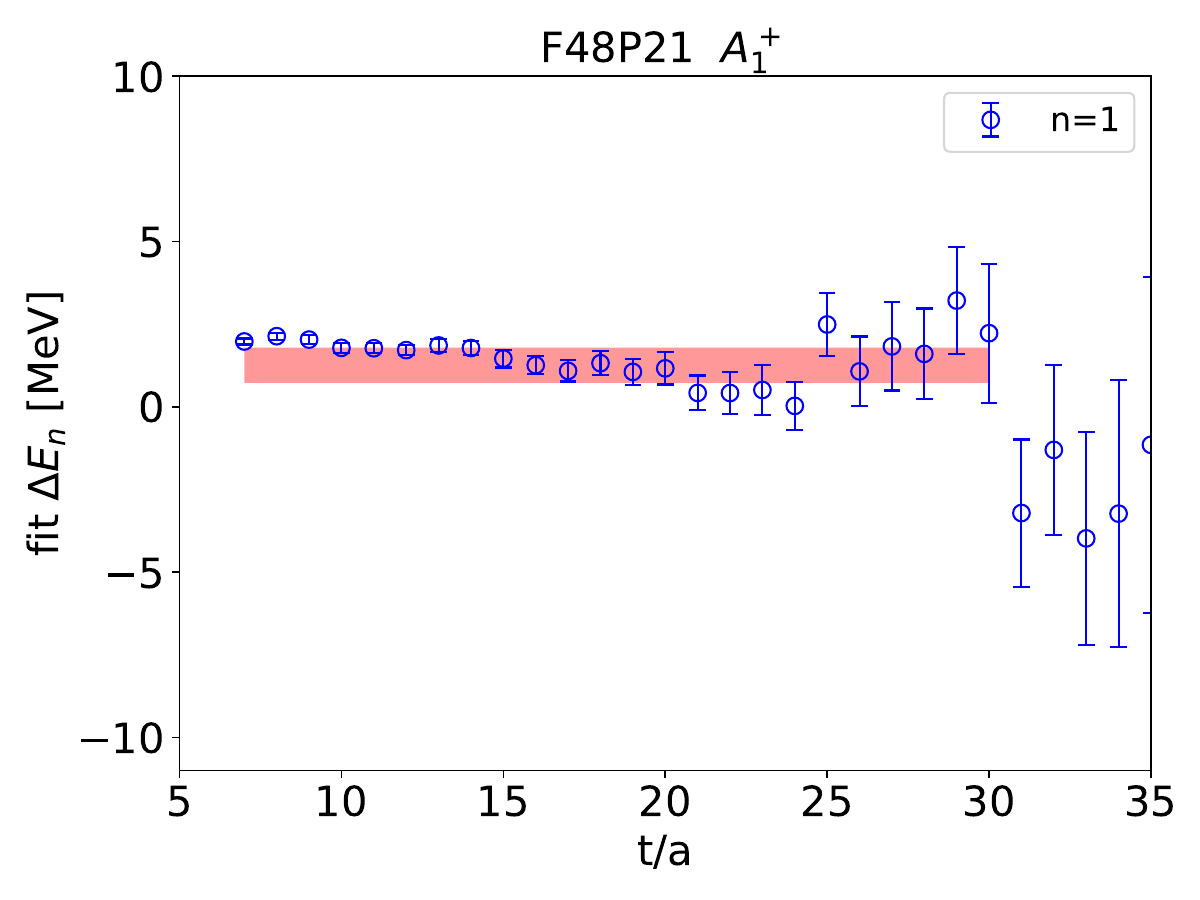}\hfill\\
\includegraphics[width=0.48\textwidth]{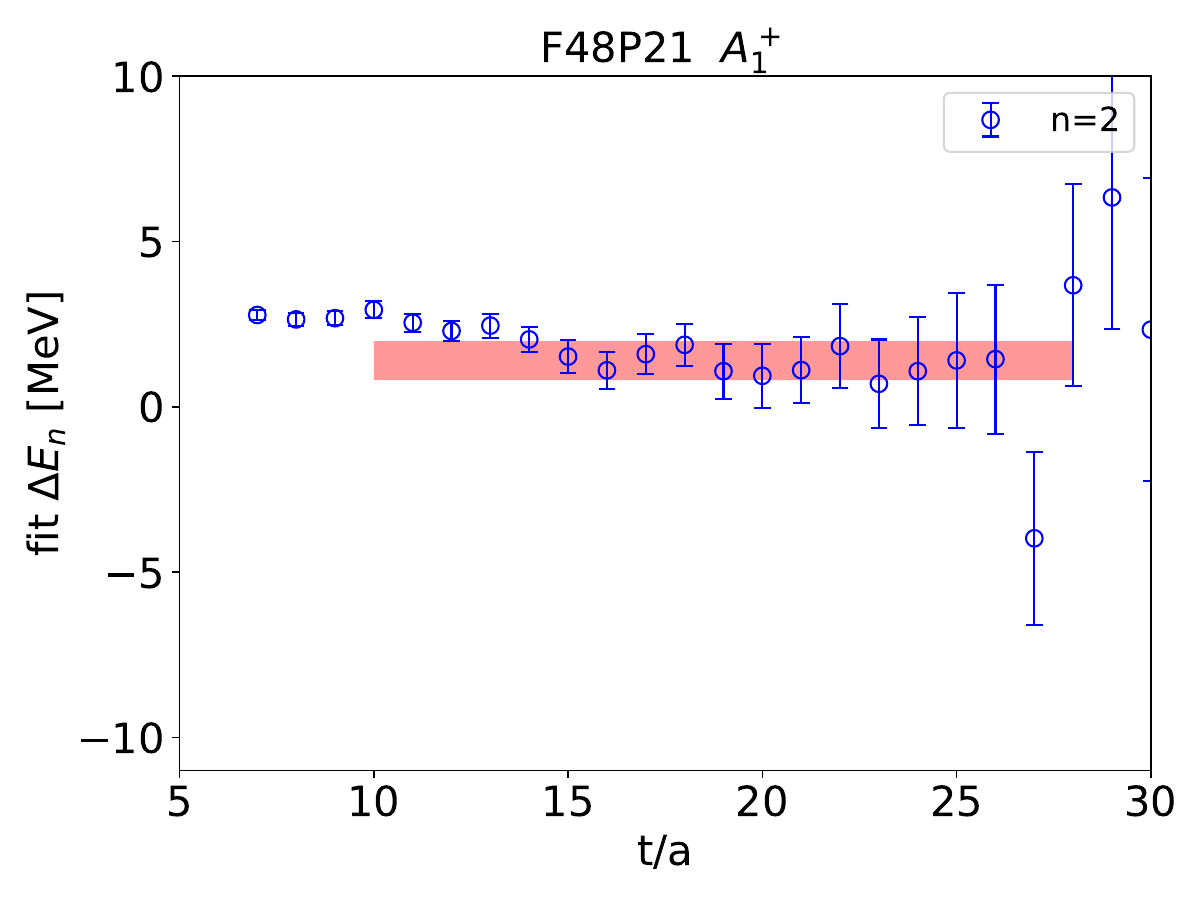} \hfill
\includegraphics[width=0.48\textwidth]{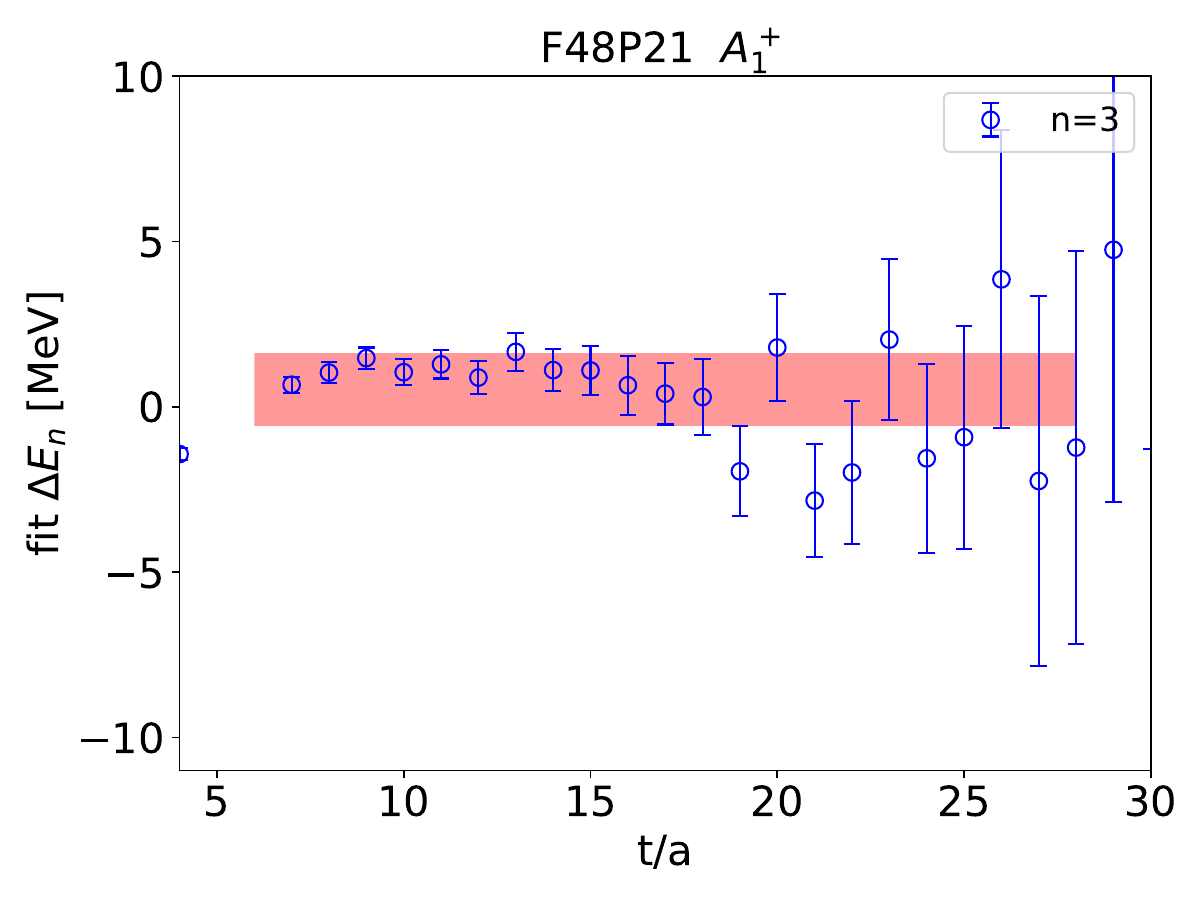}
\caption{ 
{Effective energies of $\Delta E_n$ for the $A_1^+$ irrep on the ensemble F48P21. The red horizontal bars represent the fitted values of $\Delta E_n$, with errors that include both statistical and systematic errors. The time range expanded by the bars represents the range within which various fit windows are chosen to do the fit.} \label{Fig:Fit_A1p}}
\end{figure*}

\begin{figure*}
  \centering
  \includegraphics[width=0.48\textwidth]{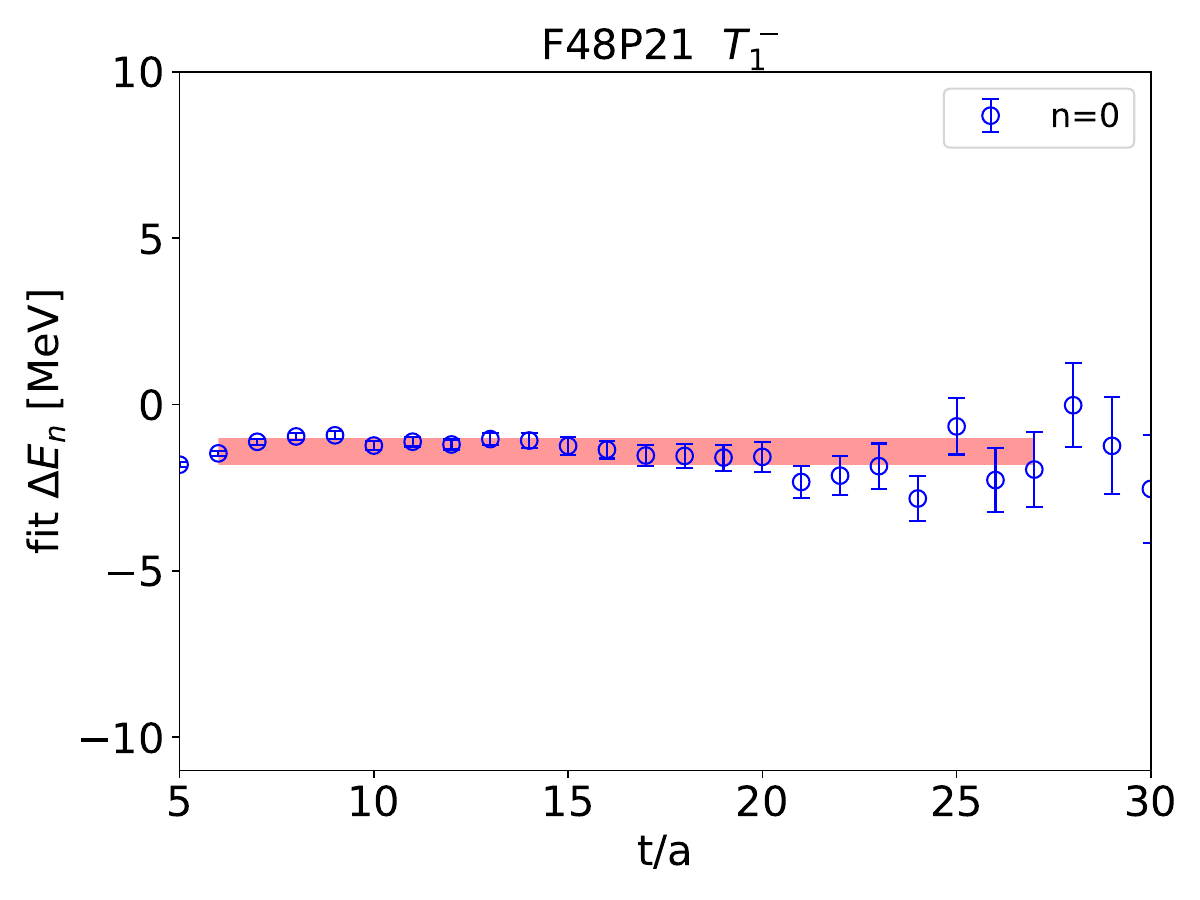}
  \includegraphics[width=0.48\textwidth]{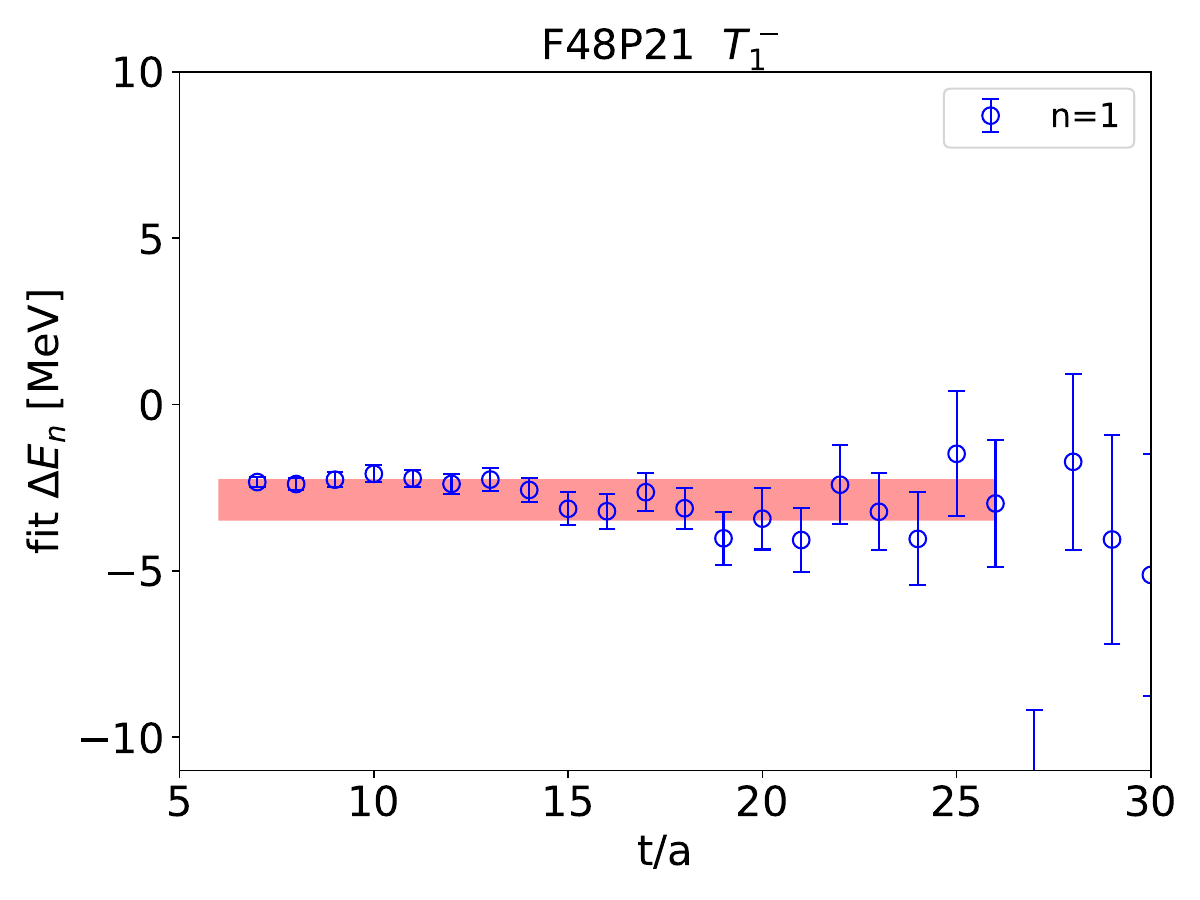}
  \includegraphics[width=0.48\textwidth]{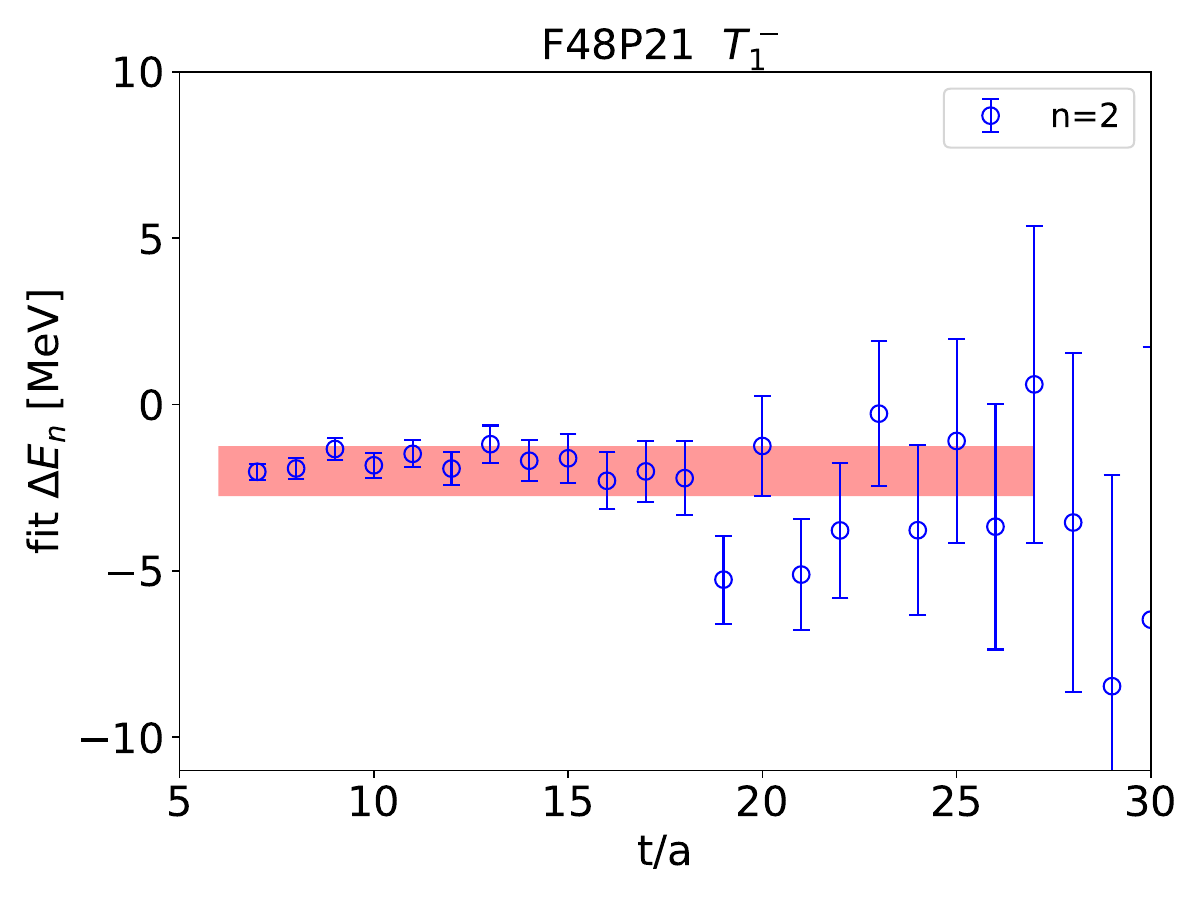}
  \caption{Same as Fig.~\ref{Fig:Fit_A1p} but for the $T_1^-$ irrep. \label{Fig:Fit_T1m}}
  \end{figure*}

After computing the correlation functions, we employ the generalized eigenvalue problem (GEVP) method~\cite{Luscher:1990ck} to determine the energy levels in each irrep. The GEVP is solved as follows:
\begin{align}
{\cal C}(t) v_n(t, t_0) = \lambda_n(t,t_0){\cal C}(t_0)v_n(t,t_0).
\end{align}
Here, $t_0$ is set to 2, and we have checked that different choices of $t_0$ do not lead to a noticeable impact on the determination of the energy levels. The energy levels are extracted from the ratio of the eigenvalues $\lambda_n(t)$ to the single-particle correlation functions
\begin{align}
R_n(t)=\frac{\lambda_n(t)}{{\cal C}_{D}^{\boldsymbol p_n}(t){\cal C}_{D}^{\boldsymbol p_n}(t)}=A^{\prime}e^{-\Delta E_nt},
\end{align}
where ${\cal C}_{D}^{\boldsymbol p_n}(t)$ is the single $D$ correlation function with momentum ${\boldsymbol p_n}$, $\Delta E_n$ is the splitting between the energies of the interacting $DD$ system and two free $D$ mesons. In the $DD$ system, where the interacting energy levels are quite close to the free energy levels, this method can partially cancel correlated statistical fluctuations and reduce the contributions of excited states by fitting the ratio $R_n(t)$, and therefore distinguish energy levels that nearly degenerate with the free ones. The interacting energies are then obtained by $E_n = \Delta E_n + E_n^{\rm free}$ with $E_n^{\rm free}$ the free energy levels. 

{To evaluate systematic errors associated with the choice of the fit window, we perform fits to each GEVP eigenvalue over all time intervals $[t_0,t_1]$ within a broad range $[t_{\text{min}},t_{\text{max}}]$, subject to the constraint $t_1-t_0\geq 6/a$. For each fit result, we assign a weight based on the Akaike information criterion  (AIC)~\cite{Akaike:1974vps,Boyle:2024grr}:  
\begin{equation}
\omega^l = \frac{1}{Z} e^{- \frac{1}{2} \text{AIC}^l},  \quad \text{AIC}^l= \chi_l^2  + 2 n ^{\text{para}} - n^l_{\text{data}}~.
\label{eq:weight}
\end{equation}
where $l$ labels the fit window, $n ^{\text{para}}$ is the number of parameters in the fit formula, which is 2 in our analysis, and $n^l_{\text{data}}$ is the number of data points in window $l$. The normalization constant $Z$ is the sum of all $\omega^l$, which makes $\omega^l$ interpretable as a probability distribution over the set of fits. This procedure is carried out for each bootstrap sample. For every bootstrap sample, we compute the weighted mean $\Delta \tilde E_n = \sum_l \omega^l \Delta E_{n,l}$. The central value and statistical uncertainty of $\Delta E_n$ are then given by the mean and standard deviation of $\Delta \tilde{E}_n $ across all bootstrap samples. 

The systematic uncertainty is estimated from the weighted standard deviation over all fit windows, evaluated on one specific bootstrap sample (the one corresponding to the mean of all configurations) :
\begin{equation}\label{sigmome}
	\sigma_{n,sys} = \sqrt{ \sum_{l}\omega^{l}_n( \Delta E_{n,l}- \Delta \tilde E_n)^2}.
\end{equation}
The total uncertainty is obtained by adding the statistical and systematic uncertainties in quadrature. 

The effective energies of the lowest four and three energy splittings $\Delta E_n$ for $A_1^+$ and $T_1^-$ irreps on the F48P21 ensemble are presented in Figs.~\ref{Fig:Fit_A1p} and \ref{Fig:Fit_T1m}, respectively. The horizontal bars indicate the central value and total uncertainty of $\Delta E_n$. The time ranges expanded by the horizontal bars represent $[t_{\text{min}},t_{\text{max}}]$. 

To estimate the systematic uncertainty in the subsequent scattering analysis (see Sec.~\ref{sec:lushcer_fromula}), we randomly select 4000 sets of energy levels from different fit windows. Taking the $A_1^+$ irrep of F48P21 as an example, four energy levels are used to determine the scattering parameters. For these four levels, the numbers of fit windows are 28, 171, 91, and 153, respectively. Thus, the total number of combinations of energy levels taken from different windows is $28\times 171\times 91 \times 153$. 
We randomly select 4000 such combinations as inputs to the scattering analysis and take the variance among the 4000 results as the systematic error arising from the choice of fit windows.
}

\begin{figure*}
\centering	
\includegraphics[width=0.49\textwidth]{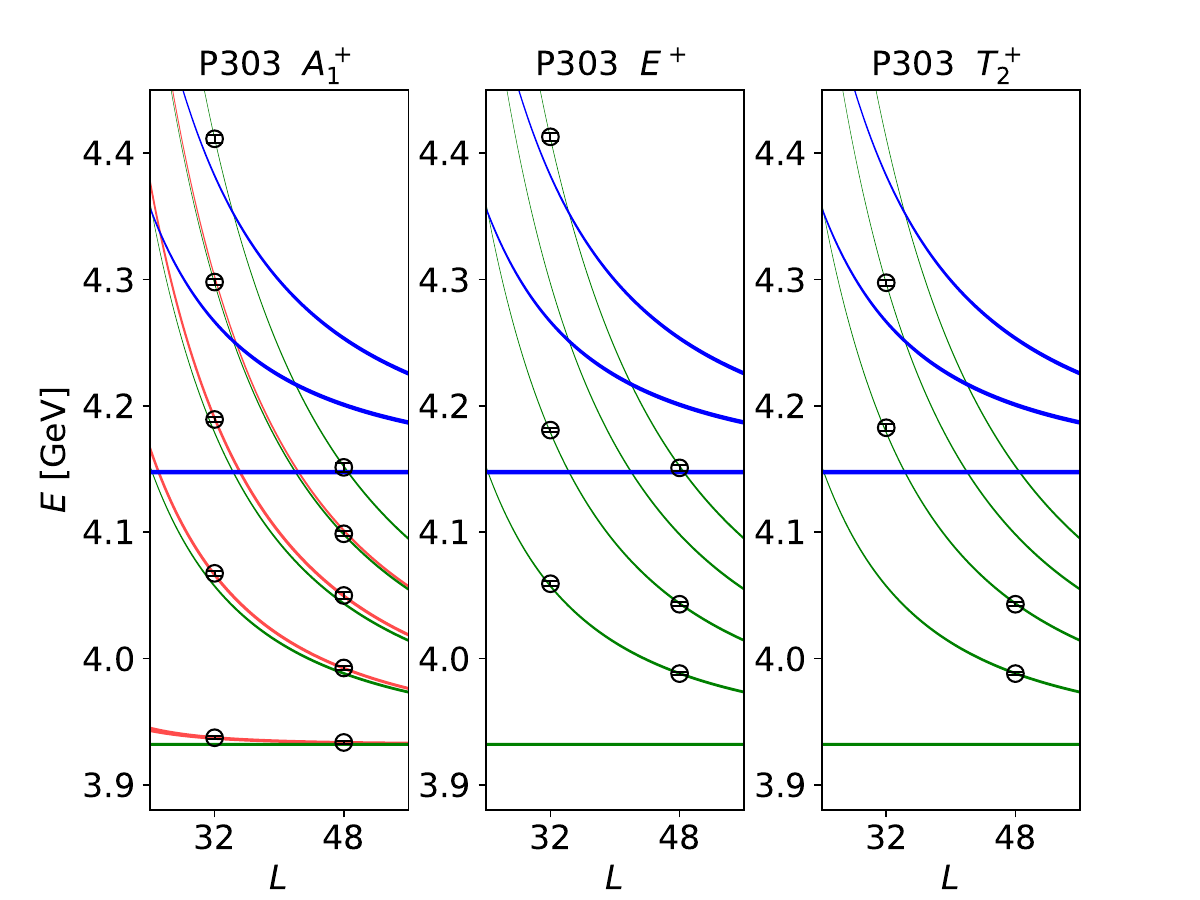} \hfill 
\includegraphics[width=0.49\textwidth]{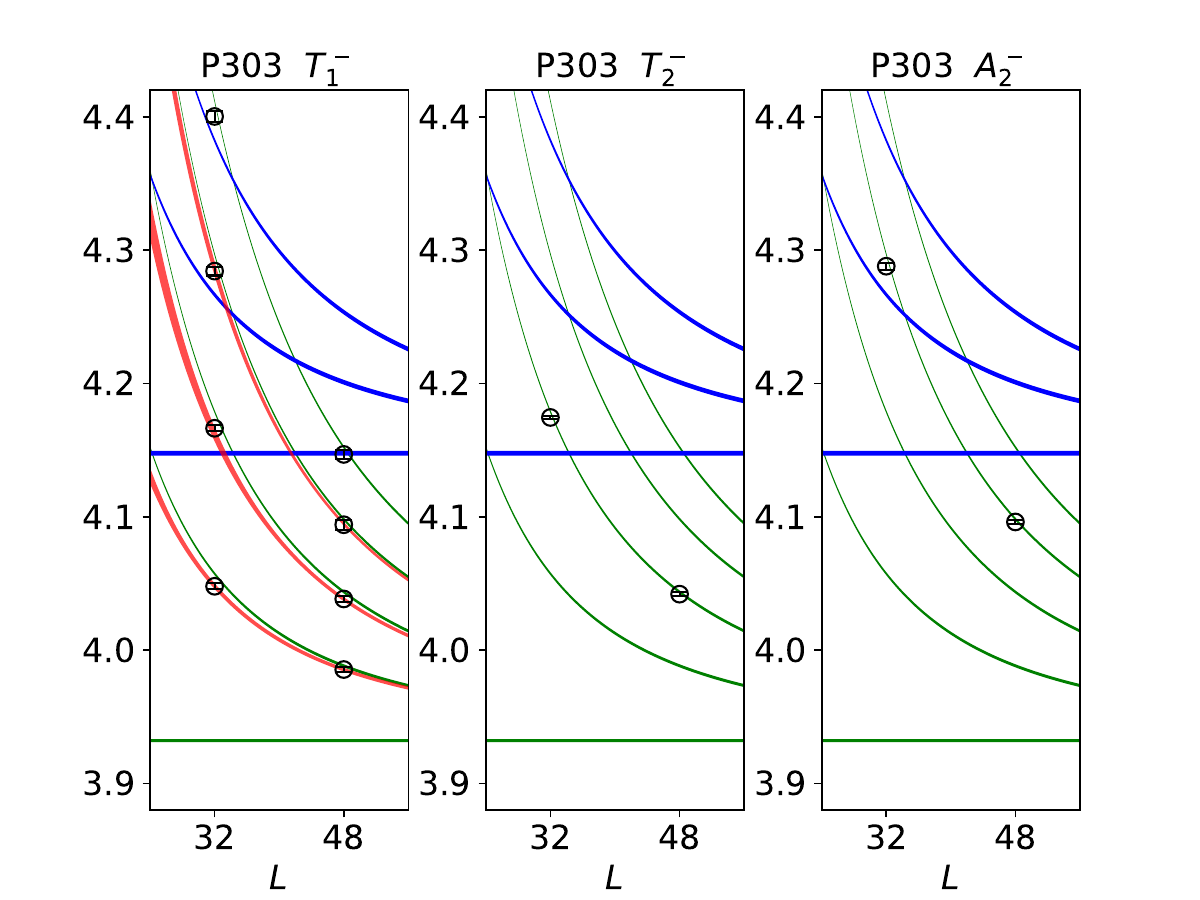}
\caption{Energy levels for the $DD$ system with a pion mass $m_{\pi}\simeq 305$ MeV. The green and blue bands represent the free energy levels for $DD$ and $D^*D^*$ systems, respectively. These free energy levels are calculated using the continuum dispersion relation (i.e., $Z=1$) with the parameters from the ensembles with lattice size $L=48$. For the $L=32$ lattice, the interacting energy levels are shifted based on the gap between the free energy levels in different finite volumes. The red bands denote the interacting energy levels fitted by the ERE parameterization.
The width of the bands reflects both {the statistical and systematic uncertainties}. \label{Fig:EL_305}}
\end{figure*}

\begin{figure*}
\centering	
\includegraphics[width=0.49\textwidth]{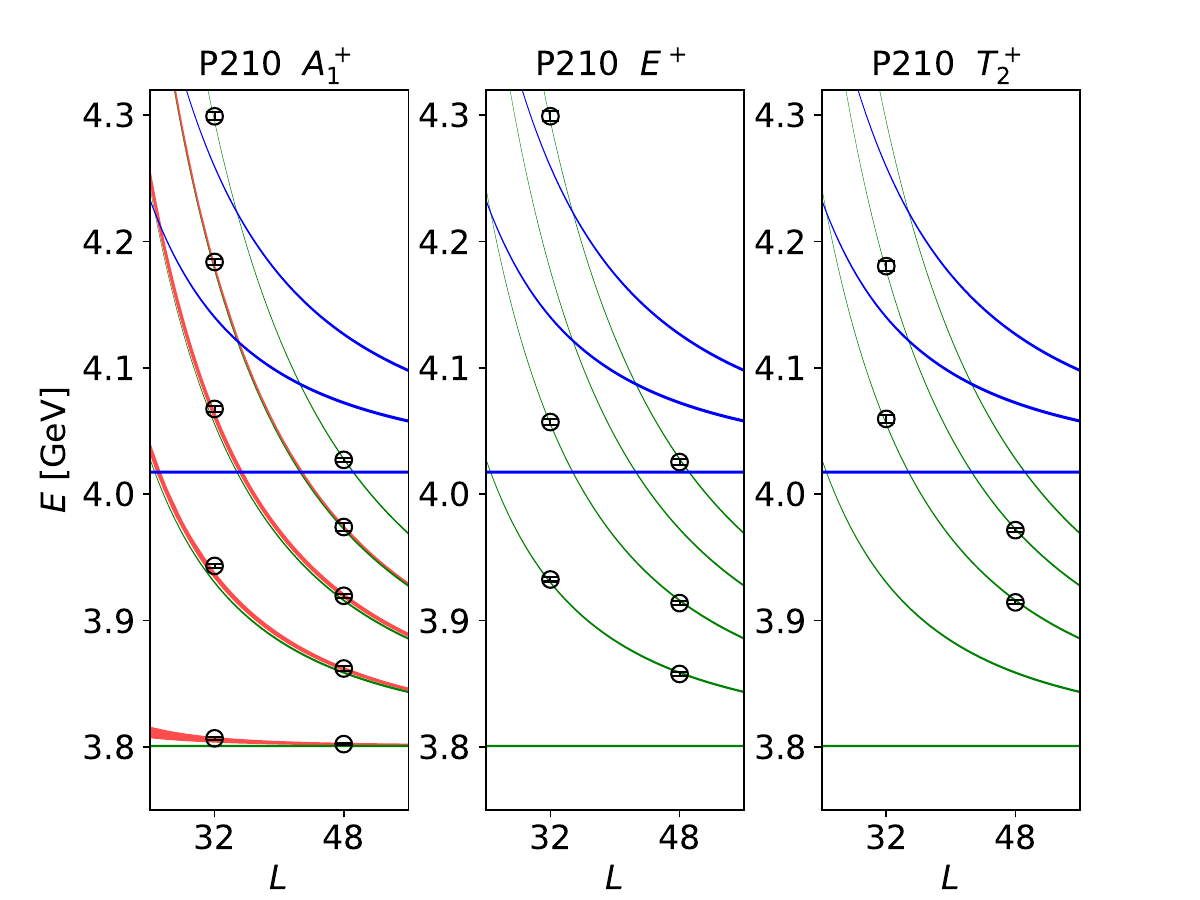}
\includegraphics[width=0.49\textwidth]{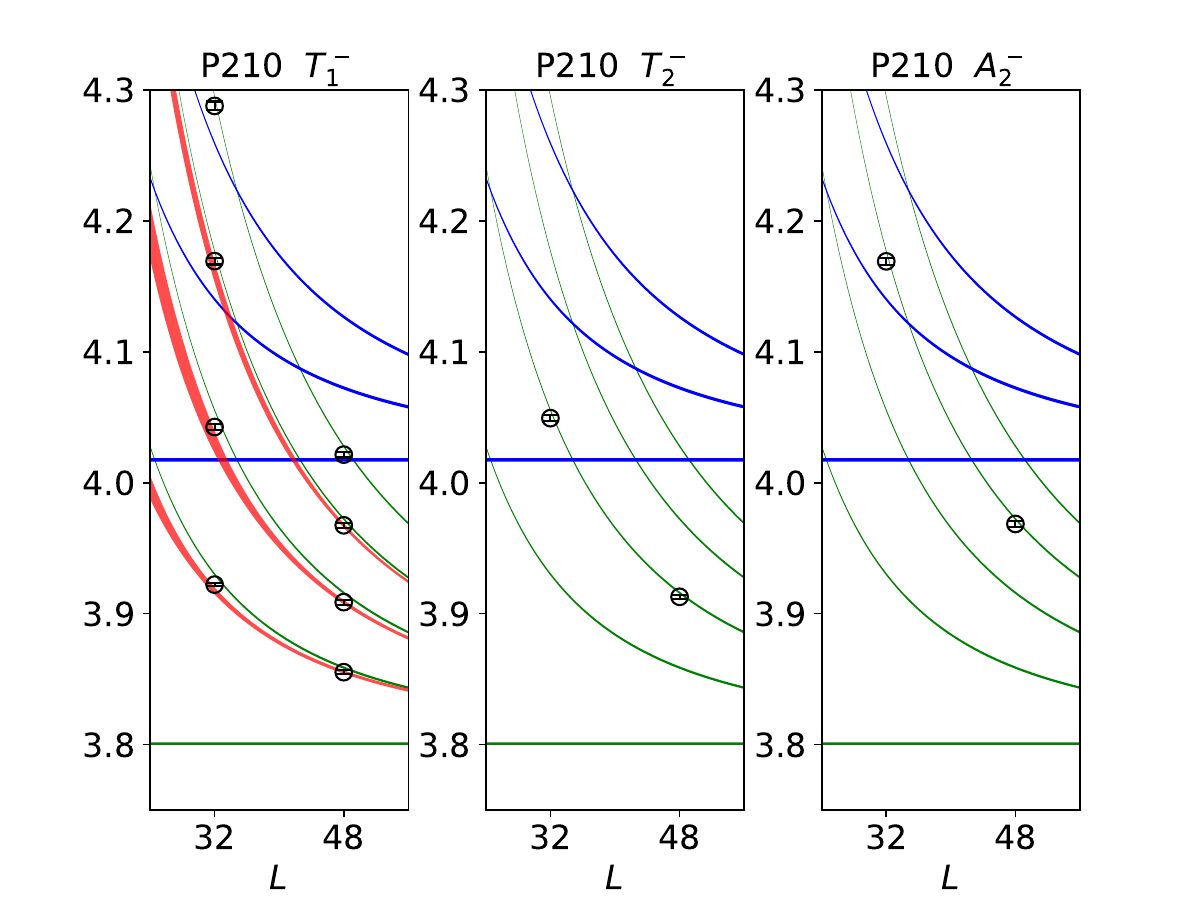}
\caption{Energy levels for the $DD$ system with pion mass $m_{\pi}\simeq 207$ MeV for the case $Z=1$. The colour bands are the same as those in Fig.~\ref{Fig:EL_305}. \label{Fig:EL_207}}
\end{figure*}

The obtained energy levels for all irreps are presented in Figs.~\ref{Fig:EL_305} and $\ref{Fig:EL_207}$. The results show that, in the $A_1^+$ irreps, the lowest energy level for each pion mass is slightly above the $DD$ threshold (see also Fig.~\ref{Fig:Fit_A1p}), indicating a weak repulsive interaction in the $S$-wave $I=1$ $DD$ system.
The energy levels in the $E^+$ and $T_2^+$ irreps almost coincide with the free ones within the current statistical precision, indicating negligible interactions in the $D$-wave and higher partial waves (the angular momenta of different irreps are listed in Table~\ref{Tab:irrep}). In the following, we will ignore the mixing from higher partial waves and utilize the energy levels below the $D^*D^*$ threshold to avoid possible contamination from $D^*D^*$, in the $A_1^+$ irrep, to determine the $S$-wave $DD$ scattering parameters in the $I=1$ channel. Based on similar observations in the $T_1^-$, $T_2^-$, and $A_2^-$ irreps, we use the energy levels in the $T_1^-$ irrep to determine the $P$-wave $DD$ scattering parameters in the $I=0$ channel.

\section{Formalism and fitting strategy}\label{sec:lushcer_fromula}

For two-meson scattering, the partial-wave phase shift $\delta_l$ is obtained from the determinant of the L\"uscher matrix,
\begin{align}
\text{det}\left(M_{lm,l^{\prime}m^{\prime}}(\widetilde p)-\delta_{ll^{\prime}}\delta_{mm^{\prime}}\text{cot}\,\delta_l\right)=0,
\label{Eq:det_LM}
\end{align}
where $\widetilde p=pL/(2\pi)$ is related to the on-shell momentum $p$. 
The energy levels are calculated within the specific little group. By selecting a particular basis for the irrep $\Lambda$, the determinant, according to Schur's lemma, reduces to~\cite{Luscher:1990ux,Gockeler:2012yj,Wu:2021xvz}
\begin{align}
\text{det}\left(M^{\Lambda}_{ln,l^{\prime}n^{\prime}}(\widetilde p)-\delta_{ll^{\prime}}\delta_{nn^{\prime}}\text{cot}\,\delta_l\right)=0,
\label{Eq:irrep_det}
\end{align}
{where $n$ and $n^\prime$ are indices indicating the occurrence of the lattice irrep $\Lambda$ in the irrep of the $SO(3)$ group labeled by the angular momentum $l$. In the cases considered in this work, the number of {occurrences} is one.}
To further simplify the L\"uscher matrix in a specific irrep, we define the function
\begin{align}
\omega_{js}(\widetilde p)=\frac{1}{\pi^{3/2}\sqrt{2j+1}}\frac{\omega_1(p)+\omega_2(p)}{Z_1\omega_2(p)+Z_2\omega_1(p)}\frac{{\cal Z}_{js}(\widetilde p)}{\widetilde p^{j+1}},
\label{Eq:omega_function}
\end{align}
where $\omega_{1,2}(p) \equiv \sqrt{m_{1,2}^2 + Z_{1,2} p^2}$, and ${\cal Z}_{js}$ is the zeta function. Compared to the original L\"uscher formula, there is an extra factor ${[\omega_1(p)+\omega_2(p)]}/{\left[Z_1\omega_2(p)+Z_2\omega_1(p)\right]}$, accounting for the deviation of the dispersion relation from the continuum one. We refer to Appendix~\ref{Sec:dsp_rlt} for detailed discussion. 

\begin{table}[tb]
\caption{\label{Tab:irrep} Irreps and the relative quantum numbers (orbital angular momentum $L$ and parity $P$) for the {$DD$ system} with $L\leq 4$ partial waves in the rest frame. {For $DD$, the positive- and negative-parity irreps correspond to isospin $I=1$ and $I=0$, respectively.}}
\renewcommand{\arraystretch}{1.2}
\begin{tabular*}{\columnwidth}{@{\extracolsep\fill}lcccccc}
\hline\hline 
 $\Lambda^{P}$   & $A_1^+$ & $A_2^-$  & $E^+$ & $T_1^-$ & $T_2^+$ & $T_2^-$ \\[3pt]     
\hline
$L$   & $0,~4$   & $3$ & $2,~4$ & $1,~3$ & $2,~4$  & $3$ \\[3pt] 
\hline\hline
\end{tabular*}
\end{table}

In this work, we focus on extracting the partial-wave phase shifts for $L=0$ and $L=1$, while the high partial-wave contributions are neglected. For the $A_1^+$ and $T_1^-$ irreps, the determinants of Eq.~\eqref{Eq:irrep_det} are reduced to
\begin{align}
-\phase_0 + \omega_{00}(\widetilde p_A)=0,\quad 
-\phase_1 + \omega_{00}(\widetilde p_T)=0,
\label{Eq:det_S_P}
\end{align}
where the coefficients of $\omega_{js}(\widetilde p)$ are taken from Ref.~\cite{Luscher:1990ux}, and the subscripts $A$ and $T$ denote the momenta for the $A_1^+$ and $T_1^-$ irreps, respectively.

Instead of translating the phase shift from interacting energy levels via Eq.~\eqref{Eq:det_S_P}, we parameterize the phase shift using the effective range expansion (ERE) and constrain the relevant parameters by fitting the energy levels in different irreps.
The ERE reads
\begin{align}
\phase_l = p^{-2l-1}\left(\frac{1}{a_{l}}+\frac{1}{2}r_{l}p^2+{\cal O}(p^4)\right),
\label{Eq:ERE}
\end{align}
where $a_l$ and $r_l$ are the scattering length and effective range for the $l$-wave, respectively. {The parameters $a_l$ and $r_l$ are determined by minimizing the $\chi^2$ defined as 
\begin{align}
\chi^2 =  \sum_{L,n,n^\prime} [ E_n(L) - E_n^{\mathrm{sol.}}(L,  a_l, r_l) ] C^{-1}_{nn^\prime} [ E_{n^\prime}(L) - E_{n^\prime}^{\mathrm{sol.}}(L, a_l, r_l) ] ,
\label{Eq:chi2}
\end{align}
where $E_n(L)$ is the $n$-th energy level obtained on the lattice with size $L$, $E_n^{\mathrm{sol.}}(L, a_l, r_l)$ is the $n$-th solution of Eq.~\eqref{Eq:det_S_P} with parameters $a_l$ and $r_l$, and $C$ is the covariance matrix of  the energies $E_n(L)$.
}
Two scenarios are considered to determine $a_l$ and $r_l$: $Z\neq 1$, where $Z$ is fixed to the values given in Table \ref{Tab:dispersion}, and $Z=1$. In each case, the interacting energy levels are calculated using the formula $E_n=\Delta E_n+2\sqrt{m_D^2+Z\bqn^2}$.
To minimize possible coupled-channel effects from $D^*D^*$, we only use the first two energy levels and four energy levels for the ensembles F32P30 and F48P30 to fit the $S$-wave scattering length $a_0$ and effective range $r_0$ in the $A_1^+$ irrep. For the $T_1^-$ irrep, the first one and three energy levels for the ensemble F32P30 and F48P30 are employed to get the corresponding $P$-wave ERE parameters, respectively. In addition, for the ensemble F48P21, the first four and three energy levels in the $A_1^+$ and $T_1^-$ irreps are used to determine the $S$- and $P$-waves ERE parameters, respectively. The energy levels for the ensemble F32P21 are neglected due to the small volume.

\section{Scattering analysis results}\label{sec:result}

The $S$- and $P$-wave ERE parameters are shown in Table~\ref{Tab:ERE_res}. In the case of $S$-wave scattering, both the scattering length $a_0$ and effective range $r_0$ are negative. In contrast, $a_1$ and $r_1$ for the $P$-wave are positive. 
Near the free energy levels, the L\"uscher's zeta function becomes highly sensitive to small variations of energy levels. Compared to the lowest interacting energy levels in $A_1^+$ irrep, the statistical uncertainties of other energy levels are enlarged by this sensitivity. Consequently, the parameters $r_0$, $a_1$, and $r_1$ have large statistical uncertainties.

\begin{table}[tb]
\caption{\label{Tab:ereparameters} ERE parameters for the $S$- and $P$-wave $DD$ scattering. The $\chi^2$ value per degree of freedom (d.o.f.) for the fits is also given. The cases for $Z\neq 1$ and $Z=1$ denote that the discrete and continuum dispersion relations are used to determine $a_l$ and $r_l$, respectively. {The first and second uncertainties for the values of $a_l$ and $r_l$ denote the statistical and systematic errors, respectively.}
\label{Tab:ERE_res}}
\renewcommand{\arraystretch}{1.2}
\begin{tabular*}{\columnwidth}{@{\extracolsep\fill}lcccc}
\hline\hline 
    & \multicolumn{2}{c}{F32P30/F48P30 ($m_{\pi}=305$ MeV)}   &  \multicolumn{2}{c}{F48P21 ($m_{\pi}=207$ MeV)}  \\[3pt]
    & $Z\neq 1$ & $Z=1$ & $Z\neq 1$ & $Z=1$\\[3pt]
\hline
     $a_0$ [fm]  & {$-0.25(6)(1)$} & {$-0.25(5){(1)}$}  & {$-0.25(5)(6)$} & {$-0.25(5){(8)}$ } 
       \\[3pt]
     $r_0$ [fm]  & {$-1.8(13)(2)$} & {$-2.0(13)(2)$} & {$-4.6(29)(11)$} & {$-4.5(30)(12)$}  \\[3pt]
     $\chi^2$/d.o.f.  & {1.28} & {1.66} & {1.02} & {1.04 }      \\[3pt]
     \hline\hline
     $a_1$ [fm$^3$]   & {$11(9)(10)$} & {$12(9){(7)}$}    & {$6(21)(14)$} & {$6(10)(8)$}   \\[3pt]
     $r_1$ [fm$^{-1}$] & {$20.2(39)(95)$} & {$20.1(38){(69)}$} & {$12.3(59)(65)$}  & {$11.9(60)(36)$}     \\[3pt]
     $\chi^2$/d.o.f.  & {1.19} & {1.23} & {1.08} & {1.03}    \\[3pt]
\hline\hline
\end{tabular*}
\end{table}

\begin{figure*}
\centering	
\includegraphics[scale=0.4]{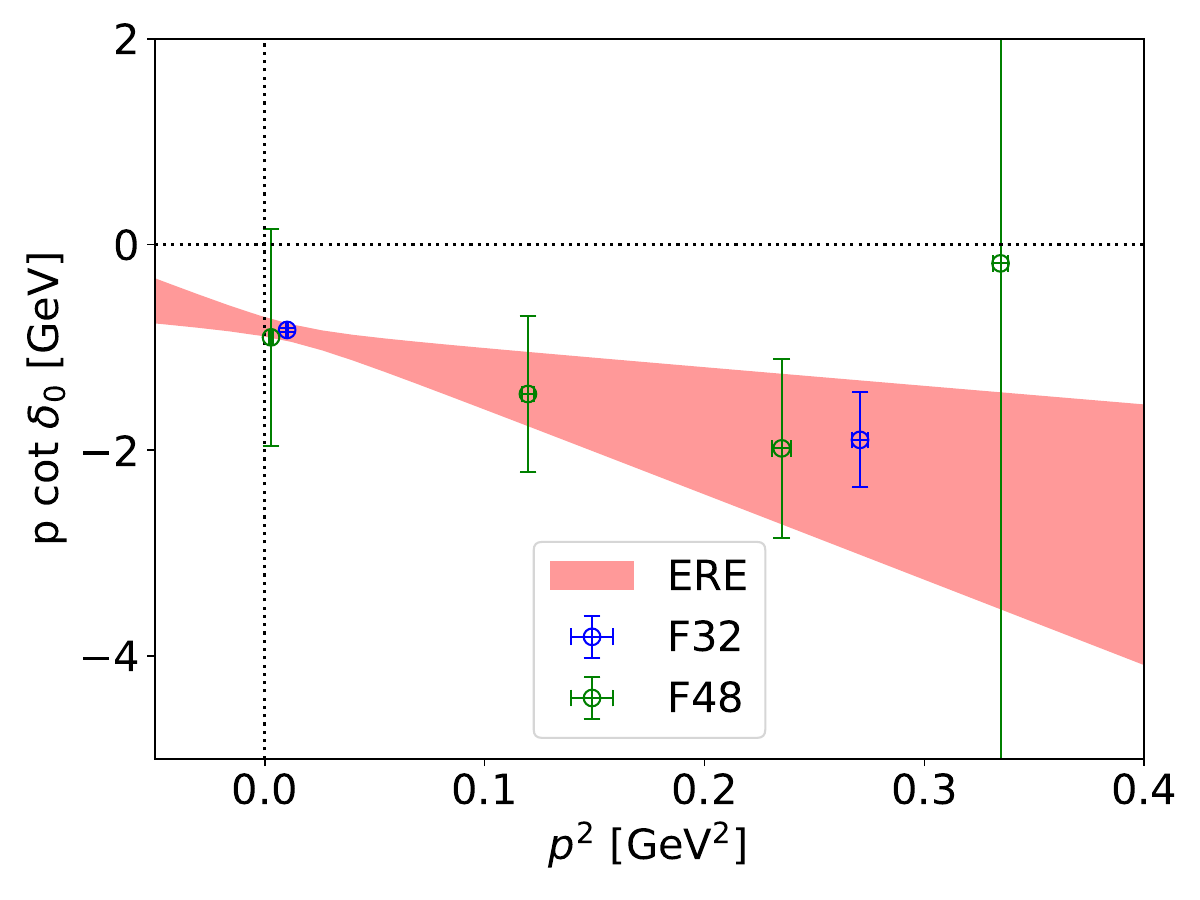}\hfill 
\includegraphics[scale=0.4]{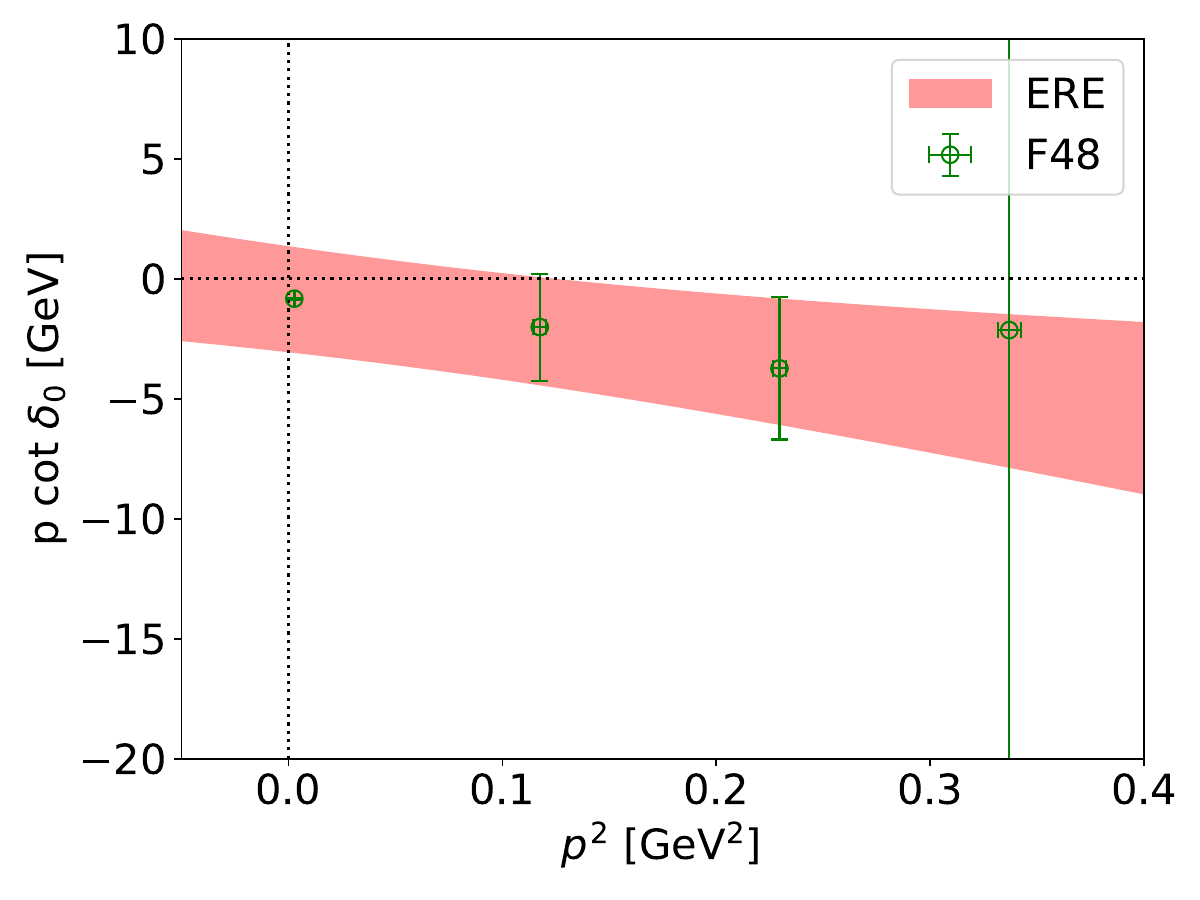}\\
\includegraphics[scale=0.4]{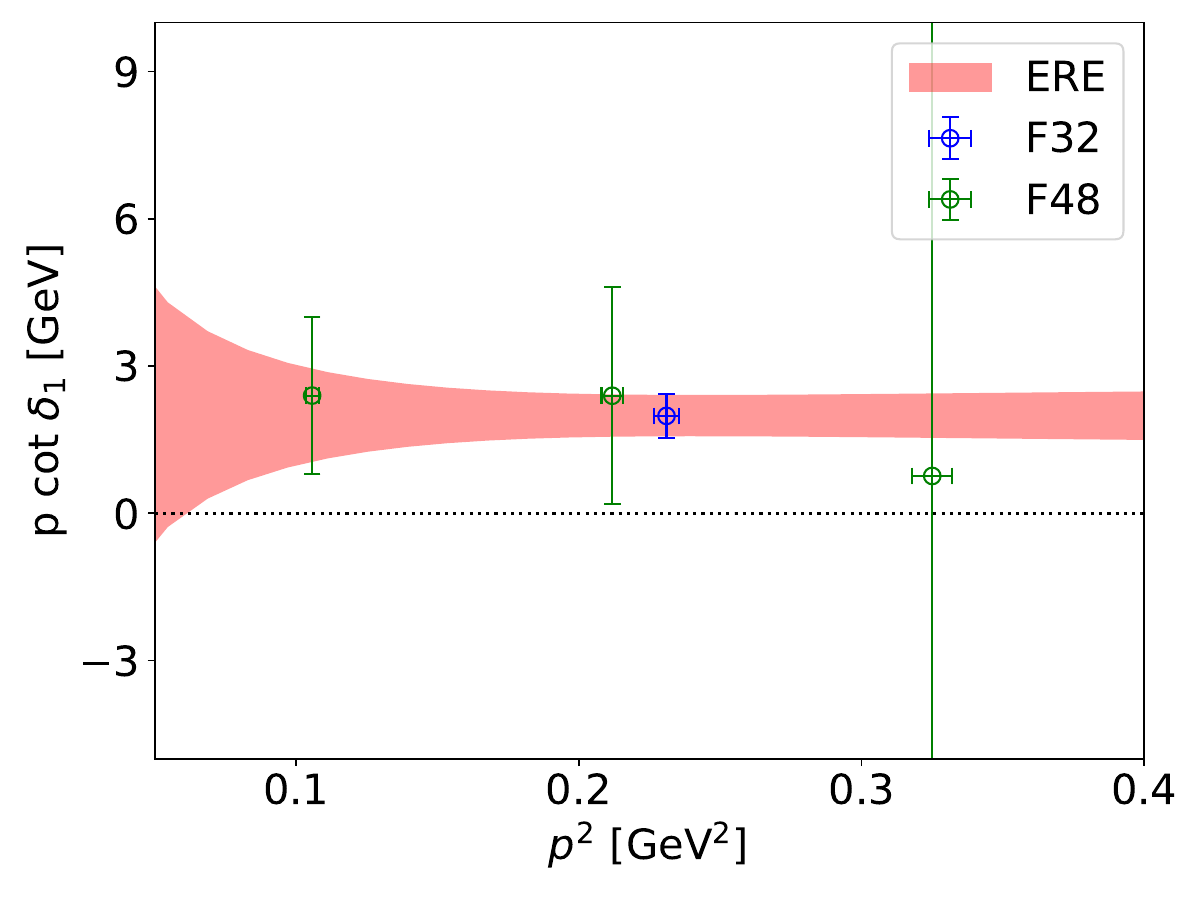}\hfill
\includegraphics[scale=0.4]{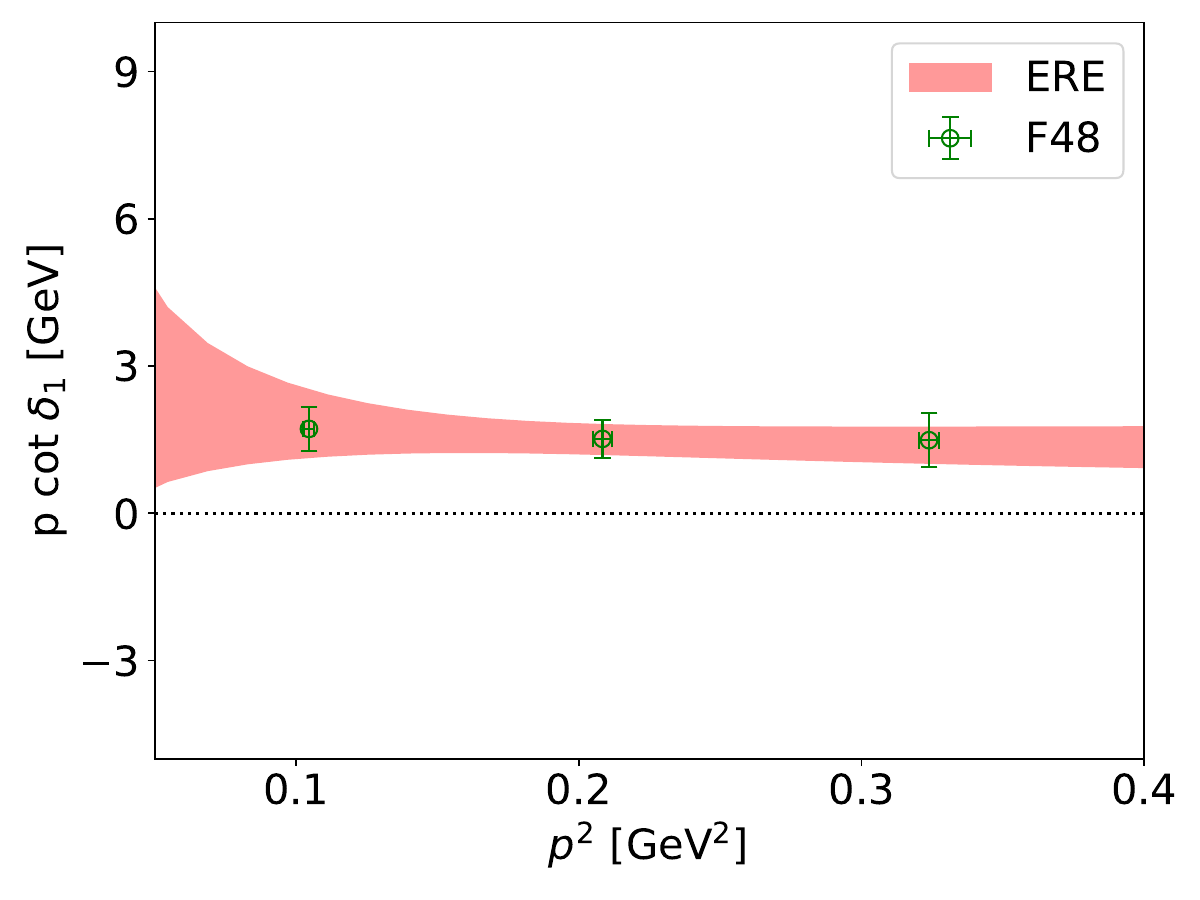}
\caption{{The variation of $p~\text{cot}\,\delta_l$ as the function of $p^2$ for $Z=1$ case. The left and right panels are the results for $m_{\pi}\simeq 305$~MeV and $207$~MeV, respectively. The red bands denote the statistical and systematic uncertainties calculated from the ERE parametrization.
} 
\label{Fig:p_cot_Delta}}
\end{figure*}

As listed in Table \ref{Tab:ERE_res}, differences in the ERE parameters between the $Z\neq 1$ and $Z=1$ cases are within $1\sigma$. {The $p\,\text{cot}\,\delta_l$ and phase shifts as a function of $p^2$ for the $I(J^P)=1(0^+)$ and $0(1^-)$ $DD$ scattering {calculated using $Z=1$} are shown in Fig.~\ref{Fig:p_cot_Delta} and Fig.~\ref{Fig:phase}, respectively.} In our convention, as the energy $E$ increases, the $S$-wave phase shift $\delta_0$ decreases, while the $P$-wave phase shift $\delta_1$ increases. The opposite signs for $\delta_0$ and $\delta_1$ indicate repulsive and attractive interactions for the $S$- and $P$-wave $DD$ systems, respectively. Considering the heavy quark flavor symmetry, the weakly attractive interaction in the $P$-wave $DD$ system is consistent with that in the $BB$ system, as calculated using a coupled-channel Born-Oppenheimer approach~\cite{Hoffmann:2024hbz}.

\begin{figure*}
\centering	
\includegraphics[scale=0.4]{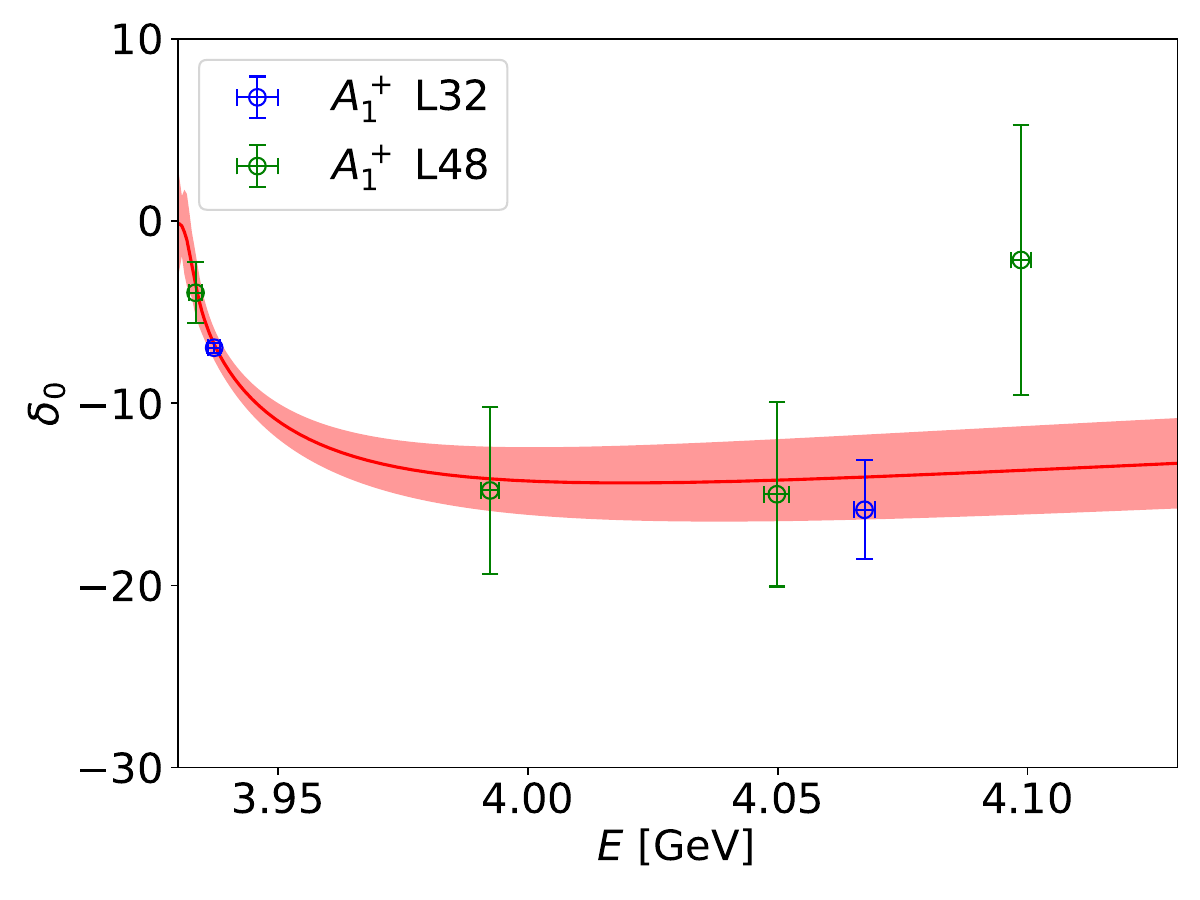}\hfill 
\includegraphics[scale=0.4]{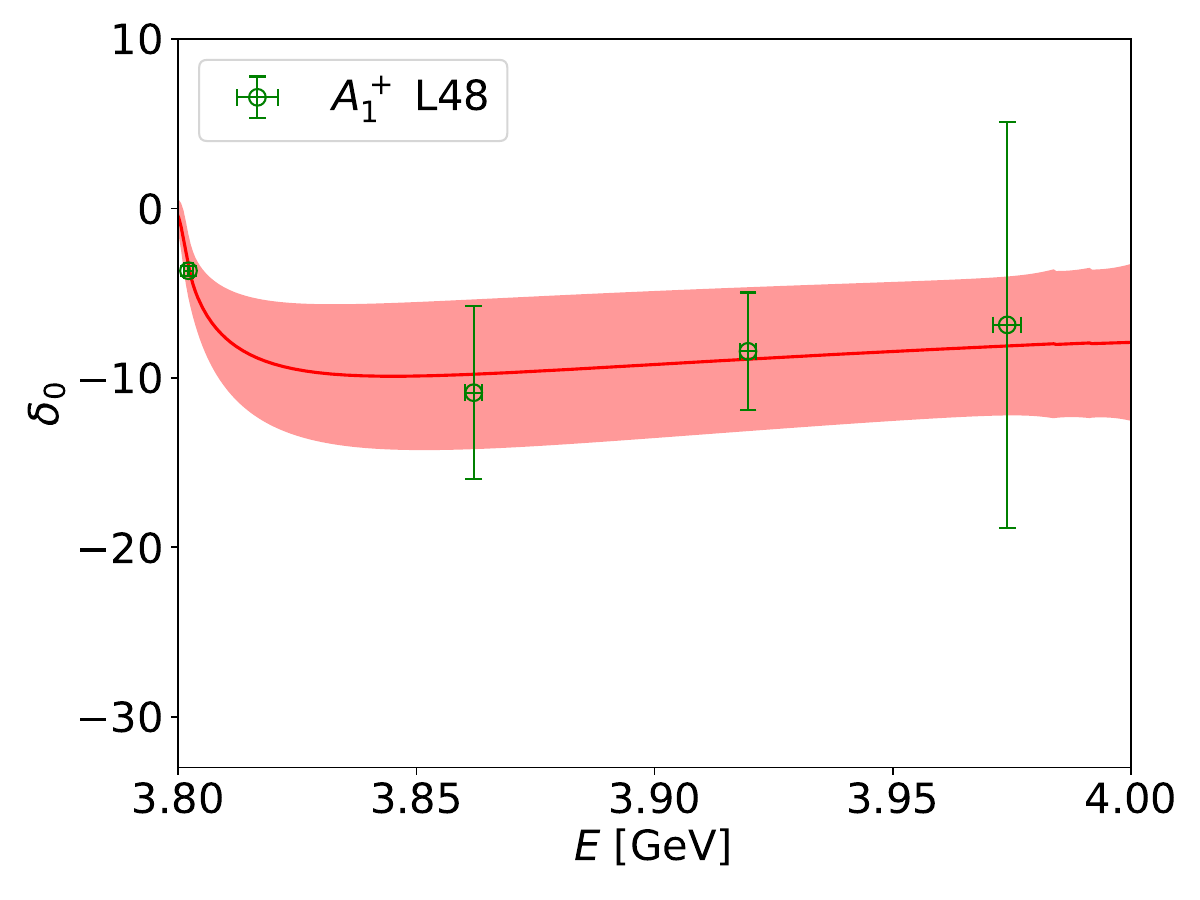}\\
\includegraphics[scale=0.4]{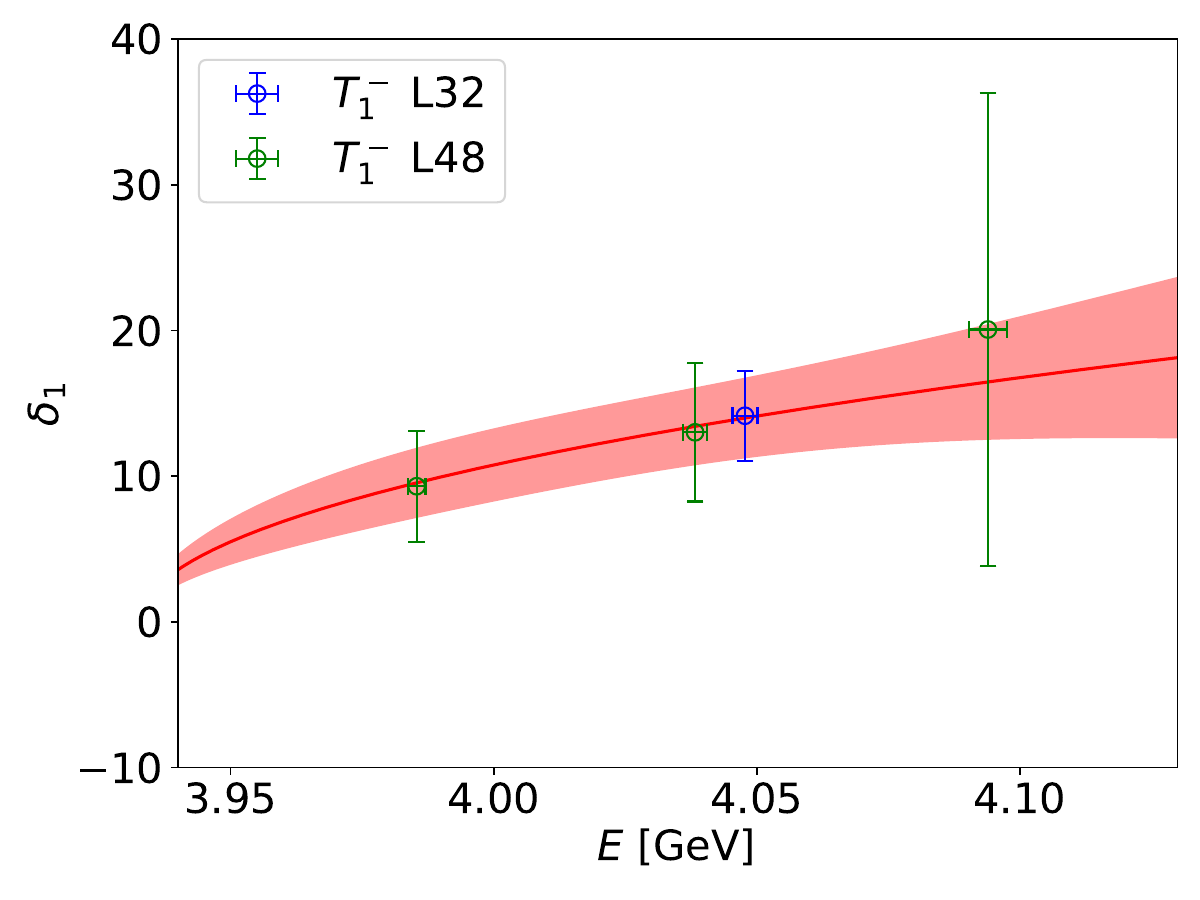}\hfill
\includegraphics[scale=0.4]{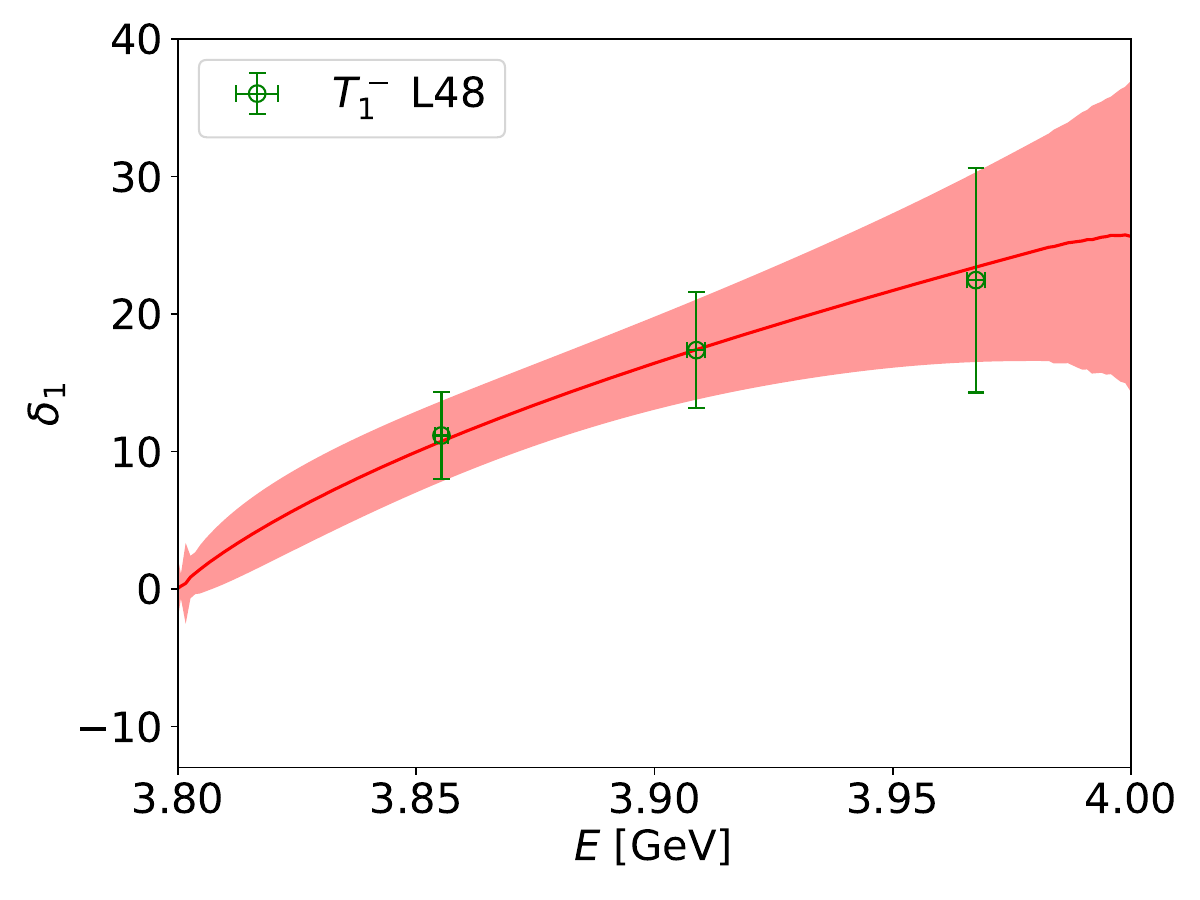}
\caption{$S$- (top row) and $P$-wave (bottom row) $DD$ scattering phase shifts for $Z=1$ case. The left and right panels are the results for $m_{\pi} \approx 305$~MeV and 207~MeV, respectively. The red lines denote the phase shifts calculated from the fits using ERE parametrization, with the bands representing the combined statistical and systematic uncertainties.
\label{Fig:phase}}
\end{figure*}

\section{Pion mass dependence}\label{sec:pion_mass}

Our results are obtained at two unphysically large pion masses. In order to determine the scattering length and effective range at the physical pion mass, we parameterize the $a_0$ and $r_0$ as functions of the pion mass and investigate their pion mass dependence for the continuum dispersion relation (the $Z=1$ case). We do not perform the extrapolation for the $P$-wave ERE parameters due to the large statistical {and systematic} uncertainties associated with them.

There are only two different pion masses 207 and 305~MeV in our calculations. Since the interaction is rather weak, it is reasonable to assume that the light-quark-mass dependence of both scattering length and effective range is perturbative. Thus, we parametrize the scattering length and effective range as simple polynomial functions:\footnote{{The formula for $a_0$ is equivalent to the relative parametrization in Ref.~\cite{Lyu:2023xro} upon neglecting the higher-order contributions, such as the ${\cal O}(m_{\pi}^4)$ term.}} 
\begin{align}
a_0(m_{\pi})=c^a_0 + c^a_1m_{\pi}^2,\quad 
r_0(m_{\pi})=c^r_0 + c^r_1m_{\pi}^2,
\label{Eq:pion_mass_dependence}
\end{align}
where the parameters $c_0^a$, $c_1^a$, $c_0^r$, and $c_1^r$ can be {solved for each sample of the fitted values of $a_0$ and $r_0$}, 
corresponding to pion masses $m_{\pi}\approx 207$ and $305$~MeV for the ensembles F32P21/F48P21 and {F32P30/F48P30}, respectively. {The resulting parameters, including statistical and systematic errors, are found to be $c_0^{a}=-0.25(10)(15)$ fm, $c_1^{a}=-0.00(6)(6)$ fm$^3$, $c_0^{r}=-6.6(56)(22)$ fm, and $c_1^{r}=1.9(25)(9)$ fm$^{3}$.}
Using the extrapolation formulae in Eq.~\eqref{Eq:pion_mass_dependence}, the $S$-wave isovector $DD$ scattering length and effective range at the {physical} pion mass are predicted to be
\begin{align}
a_0^{\text{phy}}={-0.25(8)(12)}~\mathrm{fm},\quad r_0^{\text{phy}}=-{5.7(45)(17)}~\mathrm{fm},
\end{align}
and their values in the chiral limit are
\begin{align}
a_0^{\chi}={-0.25(10)(15)}~\mathrm{fm},\quad r_0^{\chi}={-6.6(56)(22)}~\mathrm{fm}.
\end{align}
{Here, the first and second uncertainties account for the statistical and systematic errors, respectively}.

The pion mass dependence of scattering length and effective range is displayed in Fig.~\ref{Fig:extrapolation}. {The large uncertainties of the effective range prevent a solid comparison of its pion-mass dependence with that of the scattering length.} {Quantitatively, the fitted slopes $c_1^a=-0.00(6)(6)$~fm$^3$ and $c_1^r=1.9(25)(9)$~fm$^3$ are both consistent with zero (the latter at the $0.7\sigma$ level), and the effective ranges $r_0=-4.5(30)(12)$~fm at $m_\pi\simeq207$~MeV and $r_0=-2.0(13)(2)$~fm at $m_\pi\simeq305$~MeV agree within $1\sigma$; hence neither $a_0$ nor $r_0$ shows a statistically significant pion-mass dependence over the range studied.}
A more robust chiral extrapolation to the physical pion mass requires results at more pion masses in a broader range. Since we only have one lattice spacing, we are unable to extrapolate our results to the continuum limit.

\begin{figure*}
\centering	
\includegraphics[scale=0.4]{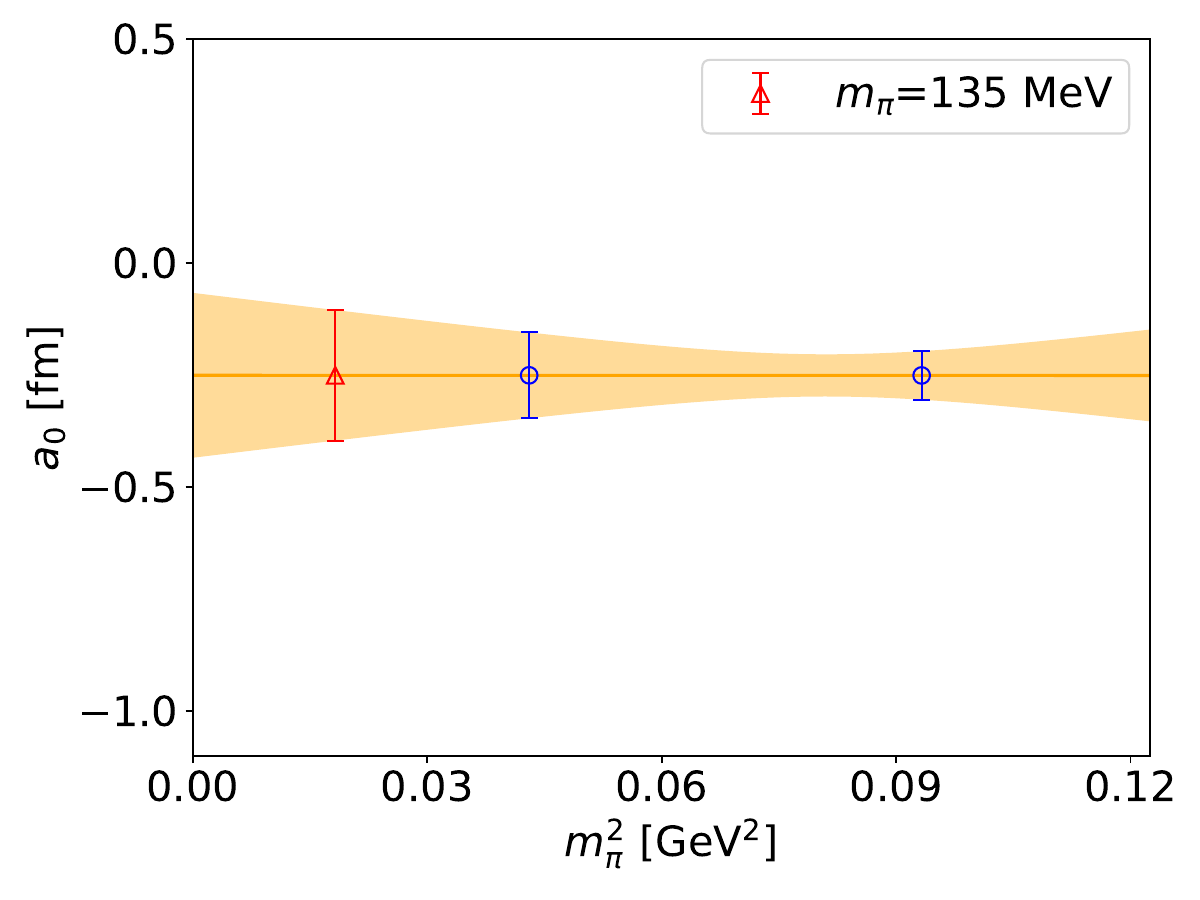}
\includegraphics[scale=0.4]{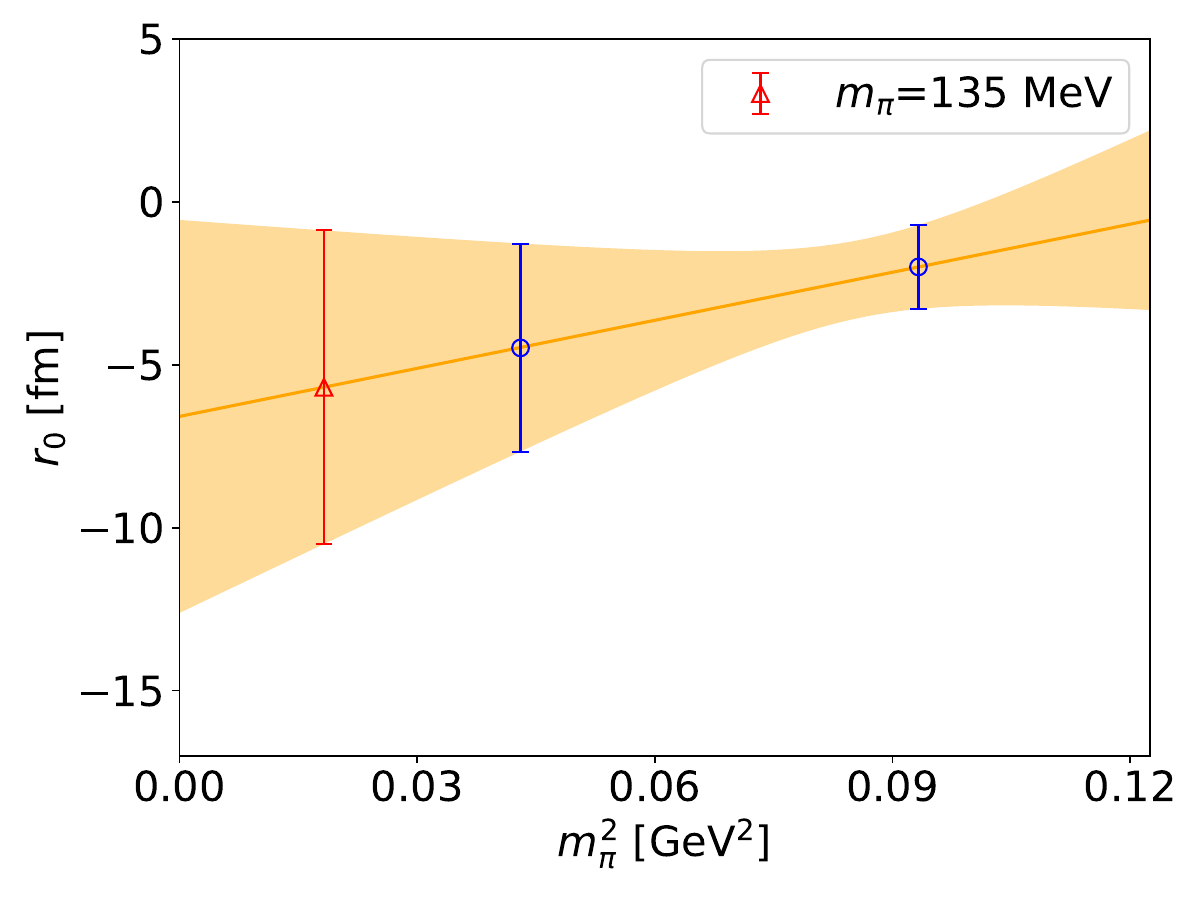}
\caption{Extrapolation of $S$-wave scattering length $a_0$ and effective range $r_0$ for $Z=1$ case. The blue bars are extracted from energy levels in $A_1^+$ irrep, and the red bars indicate the physical results, {where the uncertainties originate from both the statistical and systematic errors}.
\label{Fig:extrapolation}}
\end{figure*}

\section{Conclusions}\label{sec:summary}

Utilizing the $2+1$ flavor full QCD Wilson-Clover configurations, we calculated the $DD$ scattering in the $I(J^P)=1(0^+)$ and $0(1^-)$ channels at two different pion masses 207 and {305}~MeV. Due to the lattice artifacts, the dispersion relation of the $D$ meson on the lattice deviates from the continuum dispersion relation. We modified the original L\"uscher formula to incorporate this deviation. The $S$- and $P$-wave ERE parameters were extracted from the interacting energy levels, and the chiral extrapolation to the physical pion mass was performed. 
Our results indicate a weak repulsive interaction in the $I(J^P)=1(0^+)$ $DD$ scattering channel and a slight attractive interaction in the {$I(J^P)=0(1^-)$} channel. From a phenomenological point of view, the $S$-wave isovector $D^{(*)}D^{(*)}$ interactions from exchanging the light vector mesons $\rho$ and $\omega$ are repulsive while the isoscalar ones are attractive~\cite{Dong:2021bvy}, and recent lattice studies indeed indicate the importance of the $t$-channel light vector-meson exchange in the $DD^*$ scattering~\cite{Chen:2022vpo, Meng:2024kkp}.

{The physical $S$-wave isovector $DD$ scattering length, $a_0^{\rm phy}=-0.25(8)(12)$~fm, indicates that the near-threshold $DD$ interaction is weakly repulsive. This provides useful input for phenomenological studies of three-body systems containing a $DD$ pair. In the $DDK$ system, where binding is driven primarily by the strongly attractive $DK$ interaction and no $DD$ bound-state threshold is expected~\cite{Pang:2020pkl}, our lattice result supports treating the $DD$ interaction as a weak repulsive correction. For $T_{cc}$ and $DD\pi$ analyses, the small $DD$ scattering length gives quantitative support to approaches in which the dominant dynamics is generated by the $DD^*$ coupled channels, one-pion exchange, and the finite $D^*$ width, without introducing a separate low-energy $DD$ rescattering amplitude~\cite{Du:2021zzh,Dawid:2025wsn}. At the same time, relativistic $DD\pi$ three-body frameworks require the two-body subchannel amplitudes, including the $I=1$ $DD$ amplitude, as input~\cite{Dawid:2024dgy}. The present calculation supplies a first-principle constraint on this input and a benchmark for testing the weak-$DD$ approximation.}

In this initial investigation, we focus on single-channel $DD$ scattering in the energy range below the $D^*D^*$ threshold. In future studies, the $DD$-$D^*D^*$ coupled-channel scattering will be considered. Furthermore, lattice data with a broader range of pion masses and various lattice spacings are required to control the systematic uncertainties in a more robust way. 

\begin{acknowledgements}
We are grateful to Yi-Qi Geng for useful discussions. We thank the CLQCD collaborations for providing us with the gauge configurations~\cite{CLQCD:2023sdb}, which are generated on the HPC Cluster of ITP-CAS, the Southern Nuclear Science Computing Center (SNSC), the Siyuan-1 cluster supported by the Center for High Performance Computing at Shanghai Jiao Tong University and the Dongjiang Yuan Intelligent Computing Center. Softwares \verb|Chroma|~\cite{Edwards:2004sx} and \verb|QUDA|~\cite{Clark:2009wm, Babich:2011np, Clark:2016rdz} were used to generate the configurations and solve the perambulators. This work is partially supported by the National Key R\&D Program of China under Grant No. 2023YFA1606703; by the Chinese Academy of Sciences (CAS) under Grant No. YSBR-101; by the Generalitat Valenciana (GVA) under the PROMETEU program with Ref. CIPROM/2023/59; by MICIU/AEI/10.13039/501100011033 under grants PID2023-147458NB-C21 and CEX2023-001292-S; by the National Natural Science Foundation of China (NSFC) under Grants No.~12293060, No.~12293061, No.~12175279, No.~12125507, No.~12361141819, No.~12447101, No.~12221005 and No.~12175239; and by Guangdong Major Project of Basic and Applied Basic Research under Grant No. 2020B0301030008.
\end{acknowledgements}

\appendix

\section{Effect of the dispersion relation to L\"uscher Equation}\label{Sec:dsp_rlt}

The L\"uscher's finite volume method~\cite{Luscher:1985dn,Luscher:1986pf,Luscher:1990ux} enables us to extract the scattering information from the finite-volume spectra, which can be calculated in lattice QCD. The original L\"uscher's formula is derived by assuming the continuum dispersion relation. However, on the lattice, the continuum dispersion relation is usually broken due to lattice artifacts, especially for the heavy-flavor hadrons. To incorporate the effect of the discrete dispersion relation into the L\"uscher quantization condition, we start from the partial-wave Lippmann-Schwinger equation in the finite volume. The energy levels are given by poles of the partial-wave $T$-matrix in the finite volume with a specific lattice size $L$~\cite{Luscher:1986pf,Doring:2011vk,Doring:2012eu,Kim:2005gf,Li:2024zld}.
The finite-volume partial-wave $T$-matrix with the angular momentum $l$ ($l^{\prime}$) and its third component $m$ ($m^{\prime}$) for initial (final) states is
\begin{align}
\widetilde T_{lm,l^{\prime}m^{\prime}}(p)=V_{l}(p)\delta_{ll^{\prime}}\delta_{mm^{\prime}} + \sum_{l^{\prime\prime}m^{\prime\prime}} V_{l}(p) \widetilde G_{lm,l^{\prime\prime}m^{\prime\prime}}(p)\widetilde T_{l^{\prime\prime}m^{\prime\prime},l^{\prime}m^{\prime}}(p),
\label{Eq：T_matr_FV}
\end{align}
where $V_l(p)$ is the short-distance potential and $p$ is the on-shell momentum. 
In the finite volume, the Green's function $\widetilde G_{lm,l^{\prime}m^{\prime}}$ is
\begin{align}
\widetilde G_{lm,l^{\prime}m^{\prime}}(p)=\frac{4\pi}{L^3}\sum_{\boldsymbol q_n}^{|\bqn|<\Lambda}\left(\frac{|\bqn|}{p}\right)^{l+l^{\prime}}Y_{lm}^{*}(\hat{\boldsymbol q}_n) I(\bqn)Y_{l^{\prime}m^{\prime}}(\hat{\boldsymbol q}_n),
\label{Eq:Green_FV}
\end{align}
with a cutoff $\Lambda$ and $\hat{\boldsymbol q}_n=\bqn/|\bqn|$. Considering the dispersion relation $\omega_i(\bq)=\sqrt{m_i^2+Z_i\bq^2}$ and $\sqrt s=\omega_1(p)+\omega_2(p)$ in the center-of-mass frame, the function $I(\bqn)$ is expressed as
\begin{align}
I(\bq)=\frac{1}{4\omega_1(\bq)\omega_2(\bq)}\frac{1}{\sqrt s-(\omega_1(\bq)+\omega_2(\bq))}\approx\frac{1}{2\left[Z_1\omega_2(p)+Z_2\omega_1(p)\right]}\frac{1}{p^2-\bq^2},
\end{align}
by just keeping the positive-energy pole contribution using
\begin{align}
&\lim_{\left|\bq\right|\to p}\frac{1}{4\omega_1(\bq)\omega_2(\bq)}\frac{p^2-\bq^2}{\sqrt s-\omega_1(\bq)-\omega_2(\bq)}\nonumber\\
=&\lim_{\left|\bq\right|\to p}\frac{1}{4\omega_1(\bq)\omega_2(\bq)}\frac{2p}{\frac{Z_1 p}{\omega_1(p)}+\frac{Z_2 p}{\omega_2(p)}}=\lim_{\left|\bq\right|\to p}\frac{1}{2\left[Z_1\omega_2(p)+Z_2\omega_1(p)\right]}.
\end{align}

Using the identity for standard spherical harmonics $Y_{lm}(\hat {\boldsymbol q})$, 
\begin{align}
Y^{*}_{lm}(\hat {\boldsymbol q})Y_{l^{\prime}m^{\prime}}(\hat {\boldsymbol q})=\sum_{js}(-1)^{m^{\prime}}\sqrt{\frac{(2l+1)(2j+1)(2l^{\prime}+1)}{4\pi}}Y_{js}(\hat {\boldsymbol q})\begin{pmatrix}
l & j & l^{\prime}\\
m & s & m^{\prime}
\end{pmatrix}
\begin{pmatrix}
l & j & l^{\prime}\\
0 & 0 & 0
\end{pmatrix},
\end{align}
the finite-volume Green's function in Eq.~\eqref{Eq:Green_FV} can be reduced to
\begin{align}
\widetilde G_{lm,l^\prime m^{\prime}}(p)=\frac{4\pi}{L^3}\sum_{\bqn}^{|\bqn|<\Lambda}\sum_{j,s}\left(\frac{|\bqn|}{p}\right)^{l+l^\prime-j}\frac{1}{\sqrt{4\pi}}C_{lm,js,l^{\prime}m^\prime}\left(\frac{|\bqn|}{p}\right)^j Y_{js}(\hat{\boldsymbol q}_n)I(\bqn),
\label{Eq:Green_FV_reduce}
\end{align}
where $C_{lm,js,l^{\prime}m^\prime} $ are the Clebsch-Gordan coefficients given by
\begin{align}
C_{lm,js,l^{\prime}m^\prime}=(-1)^{m^{\prime}}\sqrt{(2l+1)(2j+1)(2l^\prime +1)}
\begin{pmatrix}
l & j & l^{\prime}\\
m & s & m^{\prime}
\end{pmatrix}
\begin{pmatrix}
l & j & l^{\prime}\\
0 & 0 & 0
\end{pmatrix}.
\end{align}
Here the factor $(|\bqn|/p)^{l+l'-j}$ can be replaced by unity, while  $(|\bqn|/p)^{j}$ is necessary so that the subtraction does not introduce a singularity at $\boldsymbol q = 0$ for $Y_{js}(\hat {\boldsymbol q})$~\cite{Kim:2005gf,Li:2024zld}.

For the single channel, the $l$-wave $T$-matrix in the infinite volume is parameterized as
\begin{align}
T_{l}(p)=\frac{8\pi \sqrt s}{p\,\text{cot}\,\delta_l-ip}.
\end{align}
The finite-volume $T$-matrix element then is
\begin{align}
\widetilde T_{lm,l^{\prime}m^{\prime}}^{-1}(p)&=T^{-1}_{lm,l^{\prime}m^{\prime}}-\left[\widetilde G_{lm,l^{\prime}m^{\prime}}(p)- \int^{|\bq|<\Lambda}\frac{d^3\bq}{(2\pi)^3}I(\bq)4\pi Y^*_{lm}(\hat{\boldsymbol{q}})Y_{l'm'}(\hat{\boldsymbol{q}})\right]\nonumber\\
&=\frac{p}{8\pi\sqrt s}\left[\delta_{ll^{\prime}}\delta_{mm^{\prime}}\text{cot}\,\delta_l-{\cal M}_{lm,l^{\prime}m^{\prime}}(p)\right],
\end{align}
where the matrix ${\cal M}_{lm,l^{\prime}m^{\prime}}$ can be expressed as the L\"uscher zeta function as follows,
\begin{align}
{\cal M}_{lm,l^{\prime}m^{\prime}}(p)\equiv\frac{1}{\pi^{3/2}}\frac{\omega_1(p)+\omega_2(p)}{Z_1\omega_2(p)+Z_2\omega_1(p)}\sum_{j,s}\left(\frac{2\pi}{Lp}\right)^{j+1}{\cal Z}_{js}\left(\left(\frac{Lp}{2\pi}\right)^2\right)C_{lm,js,l^{\prime}m^\prime}.
\label{Eq:M_define}
\end{align}
Here ${\cal Z}_{js}(q^2)$ is the L\"uscher's zeta function in the rest frame,
\begin{align}
{\cal Z}_{js}( q^2)=\sum_{\boldsymbol{n}\in \mathbb{Z}^3}\frac{\boldsymbol n^j{Y}_{js}({\hat{\boldsymbol n}})}{\boldsymbol n^2 - q^2 }.
\end{align}
The energy levels are finally associated with the infinite-volume phase shift
\begin{align}
\text{det}\,\left[\text{cot}\,\delta-{\cal M}(p)\right]=0.
\end{align}

\bibliography{refs.bib}
\end{document}